# A ferrofluid based neural network: design of an analogue associative memory

R. Palm and V. Korenivski

*Nanostructure Physics, Royal Institute of Technology, 10691 Stockholm, Sweden*

ABSTRACT: We analyse an associative memory based on a ferrofluid, consisting of a system of magnetic nano-particles suspended in a carrier fluid of variable viscosity subject to patterns of magnetic fields from an array of input and output magnetic pads. The association relies on forming patterns in the ferrofluid during a training phase, in which the magnetic dipoles are free to move and rotate to minimize the total energy of the system. Once equilibrated in energy for a given input-output magnetic field pattern-pair the particles are fully or partially immobilized by cooling the carrier liquid. Thus produced particle distributions control the memory states, which are read out magnetically using spin-valve sensors incorporated in the output pads. The actual memory consists of spin distributions that is dynamic in nature, realized only in response to the input patterns that the system has been trained for. Two training algorithms for storing multiple patterns are investigated. Using Monte Carlo simulations of the physical system we demonstrate that the device is capable of storing and recalling two sets of images, each with an accuracy approaching 100%.

## I. Introduction

We have previously [1, 2] outlined a new physical implementation of an artificial neural network (ANN) based on a ferrofluid. The device functions as an associative memory and can be used for image recognition. The capacity to store and recall a single image was demonstrated using Monte Carlo (MC) simulations of the physical system. A useful memory would of course need to store more than one image. This article investigates the issues related to multiple patterns being recognised using a device of this type, in addition to significantly improving practically all aspects of the original design of [1].

There can be no doubt that microscopic physical systems capable of mimicking the workings of a neural network would be of tremendous technological importance. Our proposed system has neuron-like elements that are about 10 nm in diameter, with a volume fraction of approximately 15%, which implies neuron densities corresponding to the whole of the human brain packed into a fraction of a cubic millimeter. Since we know very well the great capabilities of the human brain, the technological impact would be significant if we could obtain similar processing power from such a small artificial device.

Most neural network designs of today are software based or use conventional Si-based digital logic. Proposals of analogue or semi-analogue neural network implementations exist. One such proposal is based on optics [3], where the neurons are represented by light emitting diodes and photo resistors, and the synaptic matrix by a light mask. This implementation allows a precise manipulation of the synaptic strengths, thus every imaginable learning rule can be implemented. The problem lies mainly in miniaturizing the device and in the one dimensional layout of the neuron links.

One popular physical model of a massively parallell neural network is based on spin-glasses [4]. A spin-glass is a disordered array of spins, which couple through competing exchange interactions. These interactions can be excitatory or inhibitatory depending on the distance between the spins – the configuration at the core of the Spin Chip [5, 6], which uses focused laser heating to locally increase the atomic mobility in a given region of the spin-glass and thereby change the exchange coupling in the system. A strong magnetic field is then applied to write a pattern on to the spin glass film. Since individual spins cannot be moved the design relies on the ability of the system to self-organize. Using controlled heating of the entire chip the spins are allowed to move and arrange themselves in to an energetically favourable configuration. During this training phase the Chip is predicted to be capable of associating patterns [5, 6]. Large scale simulations of a chip with 4096 inputs and outputs have been performed [7], proving the capability of the device to successfully recall a large image. No simulation where more than one image have been stored in a device of spin-glass type has been reported, although some proposals for training multiple patterns exist [8]. However, the fundamental assumption that the heating required to increase the mobility of the magnetic atoms during the memorization stage (~1000 K) will preserve the inter-spin magnetic interactions responsible for the spin-glass behaviour (~1-100 K) is never justified for real physical systems. It is unclear how this problem can be solved even in principle in spin-glass based designs, which remain popular as artificial neural network models. In this paper we analyse a physical system free from the above limitation and demonstrate the associative memory function for two pattern pairs.

## II. Background

### a. Association using local energy minima

An associative memory associates two objects with each other in such a way that one of them, the value, can be obtained by providing the other, the key. The number of problems requiring this type of association is innumerable and there are many suitable software algorithms, such as hash tables, binary searches, etc. However, these algorithms are designed to work with exact information, which makes it difficult for computers to deal with things that humans do very easily, such as image recognition, where the information is diffuse. Artificial neural networks (ANN's) implemented in software or hardware have been shown to be an efficient solution for these sorts of problems.

It is now a well established fact that the synaptic weights between neurons are modified during learning. The Cognitron was one of the first artificial systems to demonstrate the capability of a system to learn and respond to stimuli without having a "teacher" instructing it on all the particulars of the training. In order for such a system to work it must be proven that it will eventually settle and not wander around endlessly. One way of doing this is to show that the algorithm employed reduces some cost function, or energy function for a physical implementation.

Hopfield [9] recognized that any physical system, which has many locally stable limit points, could potentially be used as a content addressable memory. The idea is that if the system is initialized to a state close to one of the stable states, the system will by itself converge to that state. If we in addition have a mapping for a set of patterns that should be stored onto the physical system, and this mapping put similar patterns sufficiently close to each other in the physical phase space, it would be possible to recall the full pattern by providing a very similar pattern. Now, if a procedure could be devised, which made it possible to control where the locally stable states are placed, and hence which patterns are stored, we would have a general content addressable memory, which could be used to store arbitrary patterns. Note that an associative memory could be regarded as a special case of a general content addressable memory, where the known and unknown parts of the pattern are fixed. Formally we let the system be described by N coordinates $x_1$, $x_2$, ..., $x_N$, which are components of an N-dimensional vector X. The limit points can then be labeled $X_a$, $X_b$, .... Now if the system starts out in the state $X_a + \Delta$, where $\Delta$ is some small perturbation, it will converge to the stable limit point $X_a$. Of course, if some other limit point $X_b$ exists close to $X_a$ in phase space, the perturbation would have to be very small to assure that the system will converge to $X_a$. In physical systems there is in general some noise, unless we are at absolute zero, in which case there may be more than one possible outcomes for a given set of initial values.

Hopfield proceeds to discuss a model system in which each neuron $V_i$ is either firing $V_i = 1$ or not firing $V_i = 0$. The neurons update themselves asynchronously with a mean attempt rate W according to the rule,

$$\begin{matrix} V_i \to 1 \\ V_i \to 0 \end{matrix} \; if \; \sum_{i \neq j} T_{ij} V_j \begin{matrix} >0 \\ <0 \end{matrix} \; . \qquad (1)$$

If $T_{ij}$ is symmetric this model is equivalent to the Ising model at zero temperature, since the update procedure will always lower the energy,

$$E = -\frac{1}{2} \sum_{i \neq j} T_{ij} V_i V_j \; , \qquad (2)$$

until a local minimum has been reached. In this model $T_{ij}$ is the equivalent of the synaptic strengths in a neural network. Adding noise to the system, i.e. raising the temperature, will turn the local minima into metastable states, which will be stable over certain time scales, depending on the temperature and how deep in energy the minima are.

There is also a slight modification to this scheme, in which a certain part of the system is held fixed during the entire "recall phase". This is known in the literature as constraint satisfaction. A ferrofluid of variable viscosity, discussed in the following section, is strongly influenced (constrained) by an external magnetic field, which can be highly inhomogeneous (present a pattern). The synaptic weights $T_{ij}$ in this case are the dipolar particle-particle and particle-pad interaction. These dipolar interactions are preserved at around room temperature, where the viscosity of the carrier fluid and thus the synaptic weights in the nano-magnetic network can be varied. A ferrofluid thereby appears to be an excellent candidate for an associative memory.

### b. Ferrofluids

A ferrofluid is a suspension of magnetic particles in a liquid carrier. The most striking feature of a ferrofluid is that it is a liquid with strong magnetic properties. On the microscopic scale there is a variety of interesting effects taking place due to the magnetic interactions between particles. Aggregation of particles into long chains is one result of the interactions. It should be noted that a ferrofluid despite its name is not a ferromagnet since it doesn't retain its magnetization once the field is removed, rather it should be classified as a paramagnet. The susceptibility of ferrofluids is much higher than that of ordinary paramagnets, however, which has led to the term superparamagnetism to describe the phenomenon. To avoid agglomeration due to molecular van der Waals forces, a thin layer (~1 nm) of surfactant is often applied to the particles.

Since a ferrofluid contains particles of dimensions smaller than typical domain sizes, they can be considered as single domain. The particles are often assumed to be spherical, which means the field outside such a particle is the same as that of a dipole with a magnetic moment equal to the total magnetic moment of the particle [10, p. 265]. This leads to a magnetic dipole-dipole interaction, illustrated in Fig. 1, with the interaction energy of the form

$$U_{dipole} = \frac{\mu_0}{4\pi} \frac{1}{r^3} [\boldsymbol{m_1} \cdot \boldsymbol{m_2} - 3(\boldsymbol{m_1} \cdot \hat{\boldsymbol{r}})(\boldsymbol{m_2} \cdot \hat{\boldsymbol{r}})] \quad . \tag{3}$$

One of the first theoretical studies [11] of ferrofluids was carried out in 1980 using the Monte Carlo technique developed by Metropolis et al [12]. The study agreed well with the earlier experimental results [13, 14], which had shown open loop structures with no particular orientation in zero field and long chainlike structures along the direction of the field when an external field was applied. A rather detailed model of the particles was used in the Monte Carlo simulation, dividing the hamiltonian in three terms,

$$H = U_{dipole} + U_{close} + U_{zeeman} \quad , \tag{4}$$

where $U_{close}$ is the sum of the steric repulsion and the van der Waals interaction, and $U_{zeeman}$ is the interaction with an external magnetic field,

$$U_{close} = 2Nk_B T \left( 2 - \frac{l+2}{t} \ln\left(\frac{1+t}{1+l/2}\right) - \frac{l}{t} \right) + \frac{A}{6}\left( \frac{2}{l^2+4l} + \frac{2}{(l+2)^2} + \ln\left(\frac{l^2+4l}{(l+2)^2}\right) \right) \quad . \tag{5}$$

In this formula N is the number of surfactant molecules adsorbed on the surface, $l = 2(r_{ij}/D-1)$, $t = 2\delta/D$, A is the Hamaker constant (A ≈ 25 $k_B$T), $\delta$ is the thickness of the surfactant coating and D is the particle diameter. This elaborate potential derived by Rosensweig [15] is applicable as long as the surfactant layers overlap, i.e. when l is between 0 and 2t. In our simulations we will instead adopt a simplified hard sphere repulsion potential, which often is a very good approximation [22]. The Zeeman energy governing the particle-pad interactions in our case is given by

$$U_{zeeman} = -\boldsymbol{m} \cdot \boldsymbol{B} \quad . \tag{6}$$

The size of the particles is the most important parameter determining the strength of the magnetic interactions. This is because the magnetic moment is proportional to the volume and the dipole-dipole interaction energy is proportional to the magnetic moment squared, i.e., $V \sim r^3$, $m \sim V \sim r^6$. On the other hand the characteristic particle separation increases proportional to the radius of the particles (the particle polymer shell acting as an offset only). Since the interaction strength goes as $E \sim m^2/r^3 \sim r^3$, the net effect is that the strength of the magnetic dipole interactions increase as the particle size cubed. For very large particles the gravitational pull will eventually make it very difficult to maintain a stable liquid solution. The shell thickness can be varied in preparation and has some effect on magnetic interaction strength since thicker shells increase the typical distance between the particles. The ferromagnetic material comprising the particles is also a design parameter, which through its saturation magnetization affects the magnetic moment of the nano-dipoles. It is often convenient to bundle these three parameters – the particle moment and diameter and temperature into a single dimensionles constant known as the magnetic coupling strength,

$$\lambda = \frac{\mu_0 m^2}{4\pi d^3 k_B T} \quad , \tag{7}$$

where d is the diameter of the particles including the shell thickness. The following ferrofluid parameters were used in this work: particles were taken to have the magnetic core of 15 nm in diameter and a shell thickness of 2.25 nm; the saturation magnetization of the ferromagnetic material of the particles - 1220 emu/cc; the volume concentration of particles of 15%; temperature of 300 K unless noted otherwise. The resulting coupling constant is $\lambda \approx 15$.

Long thin chains of ferrofluid nano-particles formed under the influence of an external field have been observed using optical microscopy [16]. One therefore can conclude that the chains formed in real ferrofluids can be thicker than one layer of particles. The concept of secondary particles [17 - 19] is often used to account for this behavior. The idea is that multiple primary particles are brought together in a spherical shape forming a secondary particle. It can then be shown that chains of secondary particles will attract each other if they are brought into proximity in the so called staggered arrangement, i.e. one chain is displaced by one half of a secondary particle diameter relative to the other chain in the direction of the field. If they are brought together in the parallel arrangement they will instead repel each other. This effect grows stronger with increasing secondary particle sizes. Chains tend to approach each other in the staggered arrangement even for secondary particles composed of a single primary particle. Since the distribution of the potential for these secondary particles become computationally very complicated, most simulations use single particles to approximate secondary particles. Although a model based on larger particles is more accurate the cost of doing so is a longer computation time. For single particles, thick clusters start to form at about $\lambda=4$ [17].

Up until now we have only considered mono-dispersed systems, i.e. we have assumed that all particles are of the same size. Aoshima and Satoh [20] have used cluster moving Monte Carlo techniques to investigate the properties of poly-dispersed magnetic fluids. They found that the distribution of particle sizes is very important for the structure of the aggregates formed. For large standard deviations in particle sizes, σ = 0.35, large clump-like structures start to form for weak dipole interactions. For strong dipole interactions with the same size distribution, complicated network aggregates are formed. The particles don't necessarily have to be spherical either. Walmsley et. al. [21] have simulated the micro structure of barium-ferrite dispersions, which consist of plate-like particles and is a common ingredient of magnetic recording media. It is found that stacks of particles are formed and that they influence the behaviour of the fluid.

In general we must also consider the interactions between the particles and solvent molecules, and between pairs of solvent molecules. Kalikmanov [22] has shown that this is not always necessary. Using a cell model and some reasonable approximations he was able to show that the carrier-liquid and particle subsystems can be treated as two independent subsystems as far as the thermodynamic equilibrium properties are concerned.

The relaxation of a ferrofluid when the applied field changes is goverened by two distinct mechanisms, called Brownian and Néel relaxation. The former is due to a mechanical rotation of the ferromagnetic particles and is not present when the carrier fluid is frozen. The characteristic time-scale of this is given by the Brownian rotational diffusion time, $\tau_B$ and is of a hydrodynamic origin [23],

$$\tau_B = \frac{3V \eta_0}{kT} \quad , \tag{8}$$

where V is the volume of the particle and $\eta_0$ is the viscosity of the carrier fluid. Néel relaxation occurs when the magnetization vector within the particle changes orientation. For uniaxial single domain particles the magnetization vector can be in either of two directions along the easy axis of magnetization in the absence of an external field. In the case of spherical particles the anisotropy is often due to the crystal anisotropy of the magnetic material. The anisotropy energy barrier must be overcome to change the magnetization direction. The characteristic time scale over which this occurs is given by $\tau_N$, first worked out by Néel [24],

$$\tau_N = \frac{1}{f_0} \exp\left(\frac{KV}{kT}\right) \quad , \tag{9}$$

where $f_0$ is the characteristic attempt frequency, of the order of $10^9$ Hz for typical ferromagnets [25].

Having established the energetics of a ferrofluid relevant for designing a neural network-like associative memory, we need to create suitable input and output magnetic field patterns as well as arrange a magnetic read-out. In order to understand how these tasks can be accomplished, a brief account of magnetic film inductor and sensor elements is presented in the next section.

### c. Magnetic input/output elements

The origin of magnetic fields in matter is the angular momentum of the electrons and to a lesser extent that of the nucleus. An electron is a spin ½ particle, having two contributions to the total angular momentum - spin and orbital. Without an external magnetic field the contributions from individual spins would cancel leading to a vanishing macroscopic magnetic moment. In the presence of a magnetizing field the individual spins align along the field giving rise to a paramagnetic response. This problem is treated within the framework of statistical mechanics in most introductory texts on solid state physics. There is also a contribution from an induced change in the orbital magnetic moment due to the applied field. This problem can be treated [26] by quantum mechanical perturbation theory to describe the resulting diamagnetic response. The relative strengths of these two phenomena governs the magnetic properties of most, but far from all materials.

The properties described above are applicable to all materials, but only give rise to relatively weak and non-persistent fields. Much stronger fields can be achieved with ordered arrays of dipoles which couple to each other by quantum mechanical exchange interactions. Materials which behave in this way are ferromagnets. Ferromagnetism is often treated approximately with the help of mean field theories, where the two particle exchange potential is replaced with an effective (molecular or mean) field. Such a treatment is able to many of the qualitative features of ferromagnetism.

As we know from daily experience large samples of ferromagnetic substances, e.g. iron, are not magnetized unless a field is applied. The reason for this is the formation of domains with different magnetic moments within the sample, which on a macroscopic scale cancels out the magnetic field outside the sample. When a field is applied to the sample it responds by growing the size of domains magnetically oriented along the field and for strong fields also by rotating the moments within domains to align with the field. This process is usually not fully reversible and a significant magnetization may remain even after the external field is completely removed.

For samples which are smaller than typical domain sizes the magnetic moments will always be aligned due to the exchange energy. Many samples also possess anisotropy, which aligns the spins along a preferred direction called the easy magnetization axis. In the ideal case the process of magnetization is completely irreversible and the magnetization will switch direction only when a large enough field, called the switching field, is applied along the easy axis. Such bi-stable magnetization states are used for digital storage in such devices as computer hard drives.

The above brief recourse into the basics of magnetism helps to place the discussion to follow in the proper context: the magnetic dipolar moments of the ferro-particles are intrinsically ferromagnetic with the atomic spins coupled by exchange, while the ferrofluid as a whole is paramagnetic (or superparamagnetic, referring to the 'super' spins of the nano-particles) possessing no macroscopic magnetic moment in the absence of external magnetic fields. The input and output layers (pads) are truly ferromagnetic, designed to have specific bi-stable magnetic states for writing zeros and ones to the ferrofluid layer, as described below.

There are many reasons why a ferromagnetic material can exhibit anisotropic behaviour. In this paper only anisotropy due to the shape of the input/output pads will be considered since it is a readily controlled property in lithographically defined structures. Shape anisotropy can be understood by considering the magnetic field at the surface of a ferromagnetic sample,

$$\nabla \cdot B = \nabla \cdot (H + 4\pi M) = 0. \tag{10}$$

Since M equals zero outside the sample, the divergence will be large unless the magnetization on the surface is parallel to the surface. This divergence induces a magnetic field known as the demagnetizing field, $H_d$. For a thin rectangular in-plane magnetized ferromagnetic film, the demagnetization field can be approximated as [27],

$$H_d^l = -\frac{2tM}{\pi l} \quad , \quad H_d^w = -\frac{2tM}{\pi w} \quad , \tag{11}$$

where t << w < l stands for thickness and length of the sample, respectively. The demagnetizing field will be larger in the direction along the width than along the length. This is equivalent to an effective anisotropy field, $H_k$, which is the difference between the two orthogonal in-plane directions,

$$H_k = \frac{2tM}{\pi}\left(\frac{1}{w} - \frac{1}{l}\right). \tag{12}$$

For small enough samples the spins will be uniformly distributed in the sample since domain walls of small dimension cost too much exchange energy. The upper limit on the size where this approximation is valid can be obtained by considering the additional energy due to a domain wall. An estimate considering the exchange and anisotropy energies yields a domain wall thickness $t_{dw}$ of [27],

$$t_{dw} = \pi\sqrt{\frac{A}{K}} \sim 100 \; nm, \tag{13}$$

where $A$ is the exchange stiffness constant (typically $\sim 10^{-11}$ J/m) and $K$ the anisotropy constant $K = M H_k / 2$. Thus, for film pad elements comparable or smaller than the domain wall width in the material, the limit of interest in this work, the single domain theory can be used with good results.

An external field $\boldsymbol{H}$ applied in the plane of the film produces a Zeeman contribution to the total energy in addition to the anisotropy energy. The magnetization will always lie in the plane spanned by the easy axis and the direction of the field. The total magnetic energy varies with the angle $\theta$ as,

$$U(\theta) = \frac{1}{2}MH_k \sin^2\theta - MH_x \sin\theta - MH_z \cos\theta, \tag{14}$$

where we have labelled the easy axis as z and the in-plane direction orthogonal to it as x, the configuration illustrated in Fig. 2. By minimizing this energy with respect to $\theta$ one finds that in general there will be either one or two minima [30]. In the region where there isn't a unique minimum, the spin configuration will depend on the history of the magnetization. This phenomenon is known as hysteresis, which we employ using shape-controlled ferromagnetic elements for designing the input and output pads.

In the case of an external field applied in the direction of the easy axis Eq. 14 takes the form,

$$U = \frac{1}{2}MH_k \sin^2\theta - MH_z \cos\theta. \tag{15}$$

Minimizing this energy by taking the first derivative with respect to $\theta$ gives,

$$\left(\cos\theta + \frac{H_z}{H_k}\right)\sin\theta = 0. \tag{16}$$

Eq. 16 shows that for large fields (Hz>Hk) the only solution is $\theta=0$. For small fields (Hz<Hk) the magnetic moment can point at either $\theta=0$ or $\theta=\pi$, with one of the states being a local and the other a global minimum. This gives rise to a square-shaped field excursion graph shown in Fig. 3a, where the magnetization switching occurs at the field strength equal to the anisotropy field, Hk. The hard axis M-H curve is obtained from Eq. 14 by keeping the Hx term and is shown for comparison in Fig. 3b. It does not contain hysteretic singularities.

In designing the associative memory device, we take the ferromagnetic thin films comprising the input and output pads to behave magnetically as shown in Fig. 3a, where the switching field is a design parameter we control primarily through the geometry of the pads.

## III. Device design and simulation method

### a. The proposed design

The associative memory device analyzed in this work has a ferrofluid acting as the hidden layer. Since ferrofluids are paramagnetic in nature they have no magnetic memory. Freezing the carrier fluid results in the magnetic particles being unable to move and a disordered glass-like state is obtained. Such frozen state, sometimes called a ferrosolid, has a rich phase behavior, with many local energy minima. It is also apparent that the location of these minima can be controlled by imposing an external field while the liquid is frozen. A field-controlled ferrofluid device thereby fulfills the requirements set by Hopfield for a system capable of storing and recalling patterns. Using nano-sized spin-valves highly local fields down to a few mOe in magnitude can be detected. This allows probing the magnetic landscape of the ferrofluid with high accuracy. Anisotropic ferromagnetic films with well defined hysteretic magnetic states, as those discussed in the previous section, can provide the field patterns necessary to control the ferrofluid. Thus, the design we discuss in this work has a non-magnetic plate acting as a substrate for a thin layer of ferrofluid. This substrate can be thought of as a top planarized layer of a standard microfabricated circuit made of inductive and sensor elements, which are used to control the ferrofluid and read back the results of a recall.

The processing elements, i. e., single domain magnetic nano-particles in a solution, have a magnetic moment which is fixed in magnitude but is free to rotate in three dimensions. This is in contrast to many other implementations of neural networks where the neuron states are discrete. The particles connect to each other through the anisotropic magnetic dipole interaction, which is a function of the distance between the dipoles. It is thus the position of the particles that can encode information. Due to the anisotropic nature of the dipole interaction, it can act either inhibitory or excitatory. In contrast to the proposed spin-glass chip [5-8] the sign of the inter-dipole interaction is not dependent on the dipole separation but the relative orientation of the magnetic dipoles. During the training process the particles are free to move, which changes the 'positional memory' in the device. Once the training is complete (minimum energy is reached) it is not desirable that the particles move, since that would alter the dipole-dipole interactions comprising the actual memory state. Controlling the viscosity of the carrier fluid (water, oil, paraffin, etc.) one can control how much the particles are allowed to move. The viscosity is controlled by varying the temperature of the device, in the vicinity of room temperature.

Ferrofluids become magnetized in the direction of an applied field due to the Zeeman interaction. In contrast to ferromagnets they do not retain a magnetization once the field is removed. This makes it possible to influence the positions and spin orientations of the individual particles in the ferrofluid by locally varying the external magnetic field. In our design the external field is controlled through a grid of input and output pads, composed of ferromagnetic single domain thin film elements. The binary images are encoded in the input or the output layers as a series of magnetization vectors, which for uniaxial single domain films can take only two directions (see Fig. 3a). During training both the input and output patterns are imposed in the inductive circuits incorporated into the substrate, such as the field pattern shown in Fig. 4, forcing the 'hidden' ferrofluid layer to "associate" these patterns which physically amounts to magnetic pad-particle flux linkage. During recall only the input pattern is imposed and the output pads are used to magneto-electronically read out the ferrofluid memory state.

### b. Design of input and output pads

We use a set of input and output pads to control and sense the state of the ferrofluid. The pads are composed of ferromagnetic single-domain thin film elements separated by a non-magnetic metal spacer, as illustrated in Fig. 5,6. Using the standard micro-inductor technique [33] of sending electrical currents through such tri-layer inductive elements, their magnetic state can be changed between two stable magnetization directions. For this the magnetic field of the current must be parallel to the magnetic easy axis, i.e. the current direction is perpendicular to the pad cross section shown in Fig. 5. The strong shape anisotropy makes the film elements always magnetized along the easy axis. A pad is a stack containing two thin film elements which are either magnetized parallell to each other or in the flux-closed anti-parallell state, which produces a weaker field that disappears rapidly with distance, i.e. only quadrupole and higher order terms survive.

The input pad only provides input to the system and thus does not have to probe the field. Input pads are typically in the parallell state during training and the anti-parallell state during recall. The main reason we use the flux-closed state during recall is to minimize the input-otput interference. We found previously [1] that a completely flux-closed configuration made the field too weak to successfully recall patterns. To remedy this problem the top film is made slightly thicker than the bottom film.

The output pads, shown in Fig. 6, act as both input and read-out elements. During training an output pad functions just like an input pad, but during recall it is put in the flux-closed state. Since most of the output field comes from particles near the output pad, it is important that the near fields from the output pads are kept low. This may require a more complex structure of the output pads than shown in Fig. 6, which can be the subject of a future investigation. In our simulations we will assume that the output pads produce zero field when in the flux-closed configuration.

To use the output pads as field sensors we rely on the concept of a spin valve [28]. A current flowing through the center conductor between the two magnetic films experiences the so called Giant Magneto Resistance (GMR) if the magnetic films are switched from the parallel into the anti-parallell configuration. The top film is made of a magnetically softer material than the bottom film. Using a triangle waveform for the current flowing through the conductor we can then get the top film to switch its magnetization direction first, before the bottom film switches. Depending on the orientation of the additional dipole field from the nano-particles, the period of time spent in the high resistance state will vary, as illustrated in Fig. 7. We can thus transform the local magnetic field in the ferrofluid into an electric signal suitable for electronic processing. The saturation magnetization of the ferromagnetic films in the pads is taken in this study to be 1700 emu/cc and the pad dimensions 80 x 220 nm for the input pads and 300 x 80 nm for the output pads.

**c. Simulating the ferrofluid layer**

There are three common simulation methods in statistical physics. These are molecular dynamics, Brownian dynamics and Monte Carlo methods. Molecular dynamics integrates the classical equations of motion numerically. This method is not applicable in our case since the magnetization processes cannot be described classically. Brownian dynamics uses statistical methods to sample processes which occur on shorter time scales by treating them as noise. This method has been used to simulate ferrofluids [29], where the particles translational and rotational degrees of freedom have been integrated, but the magnetization has been coupled to a thermostat. The most common methods used in the literature to simulate ferrofluids are based on Monte Carlo sampling of the Boltzmann distribution and this is the method used in this work.

In theory sampling of the canonical ensemble is very simple for a system once we know its Hamiltonian. One simply has to generate a state, calculate its Hamiltonian and weigh it by its Boltzmann factor. Simple sampling methods are seldom used in statistical physics though. This is mainly because of the huge number of states required to cover a statistically significant part of the multidimensional phase space. The concept of importance sampling, i.e. to sample states with high probability more often than states with low probability, is essential to any successful Monte Carlo simulation in statistical physics and the method used to accomplish this is often based on the Metropolis algorithm. This method has an additional advantage in that by a proper choice of parameters we can obtain the dynamics in phase space, which resembles that of the true system. This is a very important property for our simulations since we need to figure out which local minimum the system ends up in after being initialized and then left to evolve according to its own dynamics.

The Metropolis algorithm is based on the theory of Markov chains, which is a statistical process where the probabilities for how the system will evolve depend on the present state of the system, but not on its full history. After an initial transient the Markov chain will evolve into a steady state and by choosing the proper transition probabilities that steady state will be equivalent to the equilibrium distribution. To obtain the transition probabilities we assume that the system is described by a master equation such as,

$$\frac{\partial P_n}{\partial t} = -\sum_{n \neq m} W_{n \to m} P_n - W_{m \to n} P_m . \tag{17}$$

In this equation $P_n(t)$ indicates the probability of being in state n at time t. In equilibrium the probabilities are time independent so the left hand side becomes zero. If in addition to this we require that the relation is fulfilled term wise, we end up with a relation called detailed balance,

$$W_{n \to m} P_n = W_{m \to n} P_m . \tag{18}$$

Any matrix W satisfying detailed balance, will give us the correct equilibrium probability distribution P. There are a number of different choices available but the most common, and the one we will use in this work, for the Boltzmann distribution is the Metropolis choice,

$$W_{n \to m} = \exp(-\Delta E / k_B T) \text{ if } \Delta E > 0, \tag{19}$$

$$W_{n \to m} = 1 \text{ if } \Delta E < 0, \tag{20}$$

where $\Delta E$ is the energy difference between the states n and m. It is easy to verify that these relations satisfy the detailed balance condition. The method generalizes to continuous phase spaces without any problems.

Putting it all together we have the following algorithm:

1. Generate an initial state and compute its energy, including all dipolar and Zeeman interactions.

2. Make a perturbation to the system, e.g. move one particle slightly, and compute the energy difference from the previous state.

3. If the energy of the new state is lower than the previous state, accept the move as the new state. Otherwise accept the move with probability $\exp(-\Delta E/k_B T)$, if this probability is smaller than a random number uniformly generated in the range [0, 1]. Otherwise reject it and go back to the original state before the perturbation.

4. Repeat from step 2.

It is important to remember that if a state is rejected, the old state should be counted again as a part of the ensemble averages you would like to compute. Using this method results in a sequence of states which after an unknown number of steps, or Monte Carlo time, will have converged to the equilibrium distribution. It should be clear that Monte Carlo time is not equivalent to ordinary time, but is stochastic. Usually one uses steps per site as the measure of Monte Carlo time, i.e. on average each particle should be moved once per Monte Carlo time unit.

The primary reason why MC simulations of ferrofluids converge slowly using ordinary methods, is that particles are considered one by one, neglecting the collective behaviour of aggregate structures. This means that once a particle belongs to an aggregate structure, and therefore has a very low energy, it is highly unlikely that it will move to a random location. The aggregates however are also locked in position since a collective movement is not possible. The result is a collection of small non-interacting aggregates, which evolves very slowly.

A simple method to deal with this problem, which was first suggested by Satoh [31], is to allow collective movement of all particles belonging to a cluster in one trial move. A cluster is any collection of particles sufficiently close to each other. A particle belongs to a cluster if it is sufficiently close to any of the other particle in the cluster. Moving a cluster doesn't change the internal energetics of the cluster, which makes such moves much more likely to be accepted than single particle moves. This method gives a dramatic decrease in the time it takes for the system to converge.

How often should these cluster moves be performed? Obviously every move can't be a cluster move since the transition matrix for cluster moves is non-ergodic, e.g. clusters can join but they can never break up. This is a matter of some concern and in principle the mean values of interesting quantities should be evaluated using a non-cluster moving algorithm. However, if cluster-moves are only attempted sporadically the equilibrium obtained using cluster-moving will only differ slightly from that obtained using ordinary methods. To get good convergence cluster-moves only need to be attempted every one hundredth step or so, so the issue of non-ergodic transition matrix for cluster-moves can be disregarded. Fig. 8 summarizes the differences discussed above by comparing the standard single-particle and cluster-moving algorithms, with the latter producing a much higher degree of aggregation for the same MC time of 100,000 steps.

Parameter choices are very important to get good convergence for Monte Carlo simulations. The most important factor in our case is proper choice of the maximum distance a particle is allowed to move in a single step. Convergence seems to be best with maximum moves of one or a few nanometers, which is somewhat smaller than the particle diameter. It should be noted that convergence is not the only thing affected by this parameter. Since we are dealing with non-uniform fields there will be energy barriers, and the ability of particles to overcome these barriers affects which regions of phase space will be explored. This could be a problem if particles are initialized to local potential minima, resulting in them being trapped inside a region with low Zeeman energy. To avoid this we always start simulations with the maximum allowed move set to a few hundred nanometers. After a few thousand MC steps the maximum distance is reduced to a few nanometers. Cluster moves are imposed with the same restrictions as single particle moves. How much the magnetic moments are allowed to rotate in a single step doesn't affect the simulation as much. For most of our simulations this restriction is set to 0.1 radians per step in both $\theta$ and $\varphi$ directions.

### d. Modelling the pads

We need to compute the magnetic field resulting from the magnetization of the pads. Regardless of the type of magnetization it can be described as a vector field **m**(**r**), which is the magnetic moment per unit volume. If we use the field of a dipole [10, p. 246],

$$\boldsymbol{B}_{dipole} = \frac{\mu_0}{4\pi} \frac{1}{r^3} [3(\boldsymbol{m} \cdot \hat{\boldsymbol{r}})\hat{\boldsymbol{r}} - \boldsymbol{m}] \quad , \tag{21}$$

we would in general need to perform a three dimensional integration to compute the field from our pads. This integration can be done analytically and exact solutions for a number of simple shapes exists [32], but that way of computing the field from the pads is rather inflexible. A more flexible solution is to use a finite element approach and mesh the pads into single-domain magnetic cells. We use alongated thin film elements with uniform magnetization along the length of the element. By thin we mean infinitesimaly thin in theory, which in practice amounts to the thickness being much smaller than the width and length. To calculate the field from the film element we start with the field from a magnetic dipole. Using the definitions in Fig. 9 and assuming the film is magnetized along its axis with uniform magnetic moment per volume m and cross sectional area A we obtain:

$$\boldsymbol{B} = \frac{\mu_0 A m}{4\pi} \int_0^l [3(x_0 - x)\frac{\boldsymbol{r}}{r^5} - \frac{\hat{\boldsymbol{x}}}{r^3}] dx \quad . \tag{22}$$

The integral is easily solved using elementary calculus and we arrive at

$$B = \frac{\mu_0 A m}{4\pi} \left( \frac{\boldsymbol{r_l}}{r_l^3} - \frac{\boldsymbol{r_0}}{r_0^3} \right) \quad . \tag{23}$$

We see that only the endpoints matter, which corresponds to our everyday experience with macroscopic magnets having a north and south pole, and we can therefore reduce the meshing to two dimensions, which saves us a lot of time compared to the straightforward approach of integrating dipole volume elements directly. To compute the field from a film we slice up the cross section of the magnet we model into a fine grid and sum up the contributions from each part using the formula above,

$$B_{film} = \frac{\mu_0}{4\pi} \sum_{i=1}^{N} A_i m_i \left( \frac{\boldsymbol{r_{i,end}}}{r_{i,end}^3} - \frac{\boldsymbol{r_{i,begin}}}{r_{i,begin}^3} \right) \quad . \tag{24}$$

We also need to handle the recall phase. Since performing detailed simulations of the field distribution in the pads would be very difficult and time consuming, we use a simpler approach and calculate the mean field across the pads, subsequently checking whether it is above the threshold value. Thus, the status of an output bit can be in either of three states, up, down or undetermined. We will also assume that we can ignore the fields in any direction other than the easy magnetization axis of the pads, i.e. we assume that the transverse fields do not exeed the threshold value in the easy direction. In a practical device the fields from the input pads would also influence the output pads directly, but since they are known they can be compensated for, e.g. by biasing the output sensing current to produce the reverse field. We therefore disregard their influence for now. Since thermal fluctuations in the field are very pronounced it is important to average over many configurations to get accurate results. Our simulatons show standard deviations of a few Oerstedts.

### e. Quadrupole inputs

There are a number of improvements that can be made the layout used in our previous work [1]. One relates to zero link configurations, illustrated in Fig. 10, which must be done by restricting the set of patterns that can be stored. This may not be a serious problem in practice but it is restrictive from the basic design point of view. The problem lies in the artificial spatial restriction imposed on the particles through the Zeeman interaction during training. We are not using the associative properties of the ferrofluid to its full capacity. By applying the external field we are directing particles to certain areas of low energy. This is expected to limit the range of patterns formed in the ferrofluid and hence diminish the associative capacity of the ferrofluid device. For example it is almost impossible for any of the inputs, besides the four nearest neighbours that can form links, to influence an output pad. This means that patterns which are locally very similar will be difficult to store simultaneously. Usually one applies the concept of Hamming distance to describe how well separated two different patterns are from each other. In the case of binary data this is just the number of bits that differ between them. In our case it is also important that these differences are distributed evenly across the images to be stored.

By changing the input pads from dipoles to quadrupoles, as shown in Fig. 11, it is possible to always get a full set of links between the input and output pads. The quadrupoles are created by placing two identical pads next to each other, but the two will always have opposing magnetization. There could be a small distance between the pads. This completely eliminates the restriction on which patterns can be stored in the memory with the drawback of a slightly more complicated design.

### f. Training and recall procedures

We can't use any of the general training methods proposed for Boltzmann machines, since they rely on modifying individual connection strengths. Instead we will use the same principle as proposed for the Spin Chip to train our device, i.e. unsupervised learning. We impose both the input and output patterns simultaneously and let the ferrofluid settle into an equilibrium at a slightly higher temperature where the carrier is fluid. We then freeze the carrier liquid by lowering the temperature.

The device can be used in either of two modes – training and recall. During training the ferrofluid is fully mobile and the particles will respond to external stimuli by both rearranging spatially and by rotating their magnetic moments. Imposing both the input and output patterns simultaneously forces the ferrofluid into a state where the output pattern can be recalled by imposing only the input pattern. After training the fluid is frozen which results in the particles being immobilized, but the magnetic moments are still allowed to rotate. This state is known as a ferrosolid. The main idea is that the procedure used during training has positioned the particles in such a way that by imposing the input pattern the output pattern can be read out from the output pads. This property has been demonstrated for a single input/output pair in our previous work [1, 2]. Here we attempt to extend the operation to storing multiple patterns at once.

Since we can only impose one I/O pair at once some method must be used to alternate between these during training. Ultimately it would be desirable if patterns could be added, and possibly removed from the device in a sequential manner. This will not be a requirement for the training algorithms considered. Instead we will assume that we know all patterns that need to be stored on the chip from the beginning.

To store two patterns we will first use a simple algorithm that alternates between the two patterns with a fixed interval. This interval is a tunable parameter. The expectation is that the magnetic inertia of the system will result in an equilibrium being reached, which is a combination of both patterns, and which would then allow recalling them both. If the interval is too long then the old pattern will inevitably be forgotten once the new pattern is trained. On the other hand a too short interval might mean that neither pattern is adequately remembered, so finding the optimal interval is one of the tasks here.

Another possible training algorithm, which was also suggested for the spin chip, is to train the two patterns at different temperatures. The temperature affects the mobility of the particles, which makes it plausible that the ferrofluid will retain a learned memory when a new one is imposed.

With only a single pattern the training algorithm is very simple. We only need to apply the input and output pattern and let the system equillibrate, after which the system is frozen. The recall is performed by applying the input pattern only and and letting the system equillibrate, after which the average magnetic field across the output pads is sensed electrically and compared with the expected output. Using this method we were able to store one pattern pair on the device and recall it with a high accuracy [1]. The following definition was used to calculate the accuracy of the recall,

$$\alpha = \frac{1}{NT} \sum_{i=1}^{N} S_i \quad , \tag{25}$$

where $S_i = T$ if $H_i > T$ or $S_i = H_i$ if $H_i < T$, $T$ being the threshold field of the sensor, $H_i$ the average magnetic field in the easy axis direction over output pad $i$, and $N$ the number of output pads. This of course implies the field is in the correct direction, otherwise $S_i = 0$. The accuracy was found to be high for thresholds up to about 25 Oe. This accuracy definition is used throughout this work.

## IV. Results

### a. Small scale simulation

Our initial results for the alternating-pattern training procedure were quite encouraging. The training patterns were alternated during training with a period of 1000 MC steps. This means that the system never settles into equillibrium, but after a number of cycles it will approach a stable limit cycle. The initial simulations were performed on a system with 2x2 outputs and 3x3 inputs, which meant the results could be obtained quickly since only about 1500 particles were needed. The accuracy of recall was quite high up to a threshold of 10 – 15 Oe, which is within reach to be measured with the techniques previously discussed. Large variations between successive recall attempts were observed, however, probably caused by the system ending up in different local minima each time. This is shown in Fig. 12. As long as the variations are not so big that we risk retrieving the wrong value, they do not matter much in practice, since we can repeat the recall if it fails. Large variations are undesirable, however, recall attempts have to be repeated a number of times to get the full range of possible outcomes. From this perspective it would be nice if these variations could be minimized.

After a few hundred thousand MC steps the energy function settled into a limit cycle. The large energy variations in the cycle are mostly due to the time it takes for the particles to orient themselves with the field, and not so much because of particle movements. The recall results were obtained using the configuration frozen after a complete cycle. In theory recall results may vary depending on where in the cycle the ferrofluid is frozen, but this has not been systematically investigated. Compared to the single pattern results [1] the configuration of particles in the two-pattern case is more disperse (see Fig. 13), i.e. the links are not so rigid because the particles never completly settle. This may be a part of the reason why the recall results vary relatively much. Looking at the recall results in Fig. 12 there can be no doubt that the system has "memorized" both patterns. However in a single recall attempt the pattern cannot be recalled with a hundred percent certainty. There are instances when the field will fluctuate over 10 Oe in the wrong direction leading to a misreading of that bit, but on average all bits will be read correctly.

### b. Simulation of a large system

Because of the expected raher local behaviour of the device (dipolar fields decay as $r^{-3}$) the move to a larger system may seem unneccessary, however as we shall see there are important differences and lessons to be learned. A problem with the larger system is that much more particles are needed and hence the simulation time increases substantially. By using a cutoff for the dipole interaction the time will only increase linearly with the number of particles. We set the cutoff at about 10 particle diameters for our simulations. We will investigate a system with 6x6 inputs and 5x5 outputs, containing roughly 6000 particles.

The encouraging results for the small system is a good motivation to test the same procedure on the larger system. The patterns we will attempt to store and recall are shown in Fig. 14. We have purposely avoided any conflict between the patterns, i. e., there is always at least one of the four closest input bits that changes for every output bit. This aspect is detailed in Fig. 15. The only possible problem bit is the one where all the input bits change, but the output bit remains unchanged. Due to the symmetry of the system this in theory is equivalent to the case where none of the 4 input bits changed, but the output bit did. Successful recall of this bit would thereby indicate that the system isn't entirely local and that the interctions beyond the nearest neighbours are significant.

For this configuration the recall accuracy of both patterns is only 60% at 10 Oe threshold. With such low accuracy it is time for a radical change of design. One source of error could be that the output pads are too close to each other. This problem can be addressed by using the layout illustrated Fig. 16. To eliminate the variations in successive recall attempts we pack the pads tighter, which increases the field strengths. The cell size after this design change is only 300 nm squared. Another advantage of this more compact layout is that much fewer particles are needed. In the simulations to follow only about 2300 particles are used for a system with 6x6 inputs and 5x5 outputs, which means that the number of particles per cell is just 1/3 of the number used in the other design. This significantly speeds up the simulations but also results in a somewhat lower storage capacity of the device.

The recall results improve significantly compared to the previous layout. The recall is biased in favour of pattern 2 for the simple reason that it is the last pattern in the cycle to be trained. The system was trained with a switching period of 1000 MC steps for a total of 1000000 steps. After every 100000 steps the recall accuracy was probed. The results are shown in Fig. 17. The best results show almost perfect recall of pattern 2 and up to 80% accuracy for pattern 1. There is no indication that the accuracy would rise if the training was allowed to go on further. It is highly probable that recall accuracies of 90% for both patterns can be obtained if training is stopped at a neutral point in the training cycle.

A new training method was also tried. Instead of alternating between patterns, which is a rather contrived way to train the device since it requires a complete knowledge of all patterns to be stored, and further doesn't scale to more patterns in a sensible way, we train the patterns successively. First the device is trained with pattern 1 until it is close to equilibrium, then pattern 2 is applied for a short while before the device is frozen. If successive memories are remembered at slightly lower temperature this will make certain structures in the ferrofluid more stable and thereby the previous memories will not so easily be forgotten. To test this principle we exagerrated this idea and stored the first memory at 600 K and the second at 300 K. Using this technique we could get 90% - 95% accuracy for both patterns at ~10 Oe threshold (Fig. 18). The peak accuracy for the combined result (not shown) occurs after about 5000 MC steps of training with the second pattern. Letting the training run longer than that increases the accuracy of pattern 2 to 100%, while lowering the accuracy of pattern 1. The reverse is true if the training of pattern 2 goes on for a shorter time. The recall was performed at 300 K. A comparison of the particle distributions for the system after only one pattern was trained and after the second pattern was optimally trained is shown in Fig. 19 and illuminates the principle of association used in the proposed artificial neaural network – particle-particle and particle-pad magnetic flux linkage from intup to output controlled by the varyale viscosity of the ferrofluid carrier.

This sequential (rather than alternating) training procedure was investigated further with the second pattern being stored at an even lower temperatures, 50 K, 60 K and 75 K. At such low temperatures the dipole coupling strength is much larger than normal (factor ~6). This meant that the first pattern was never completely forgotten. Even after 500000 MC steps the accuracy of this pattern was over 90%. The second pattern also reached accuracies over 90%, but as of now we have not managed to achieve 100% accuracy for both patterns simultaneously.

The potential problem bit due to locality issues mentioned above was in fact recalled correctly when patterns were stored at different temperatures, but not when the alternating pattern storage procedure was used. This is encouraging but it is too early to speculate whether this really is due to the input pads being further apart.

An illustrative summary of the results obtained in this work is shown in Fig. 20, where two pattern pairs with the respective field maps in the ferrofluid layer as well as the corresponding recall accuracies are compared for the two training methods studied – alternating-pattern and variable temperature training. Such multiple field maps due to the local spin distributions are the actual, input-activated memory states of the proposed associative memory device.

## V. Conclusions and outlook

Developing an associative memory based on ferrofluids will be a challenge. The results in this paper should be interpreted keeping in mind that this investigation has only just begun and that the simple devices examined here in no way do justice to what one can expect from a real ferrofluid based associative memory. On the other hand the device we propose is not merely a theoretical speculation, but one that can be built using present day technology. The performance of the device is limited by a number of factors, such as the ferrofluid properties, pad layout, and training algorithms. Improvements in all these areas are probably necessary before a really useful device can be constructed.

In theoretical models, such as the Hopfield model, the interconnections between processing elements are not restricted significantly by spatial constraints. In a ferrofluid the dipole interactions decay as $1/r^3$ and even faster for spin-glasses. This means that the number of effective interconnections between the processing elements is limited in the proposed design, which is true for any physical (non-biological) system. These restrictions are expected to decrease the memory performance of designs based on real physical systems as compared to ideal theoretical models. How much these restrictions affect the storage capacity is not entirely known. Nevertheless if some methods could be devised to reduce the rather fast decay of the interactions, it could only help the performance. A tempting solution is to make very elongated particles whose ends would then effectively act as monopoles, so that the decay of the strength of the 'inter-neuron' connections would reduce to only $1/r$, resulting in a significantly extended coupling radius. The problem with this solution is that elongated particles possess shape anisotropy and therefore require a high switching field to rotate while in the ferrosolid phase.

The move to three dimensional devices will open up a whole new world of possibilities for complex chain formations. In 2D chain formation is limited by particles not being able to move across the 2D plane. This is the same problem that inhibits 2D dynamical systems from showing chaotic behaviour. With a thick ferrofluid layer instead of a monolayer many new design possibilities would open up.

It is clear that pattern formation in current designs is limited by the paths provided by the Zeeman energy from the pads. New pad layouts could improve this and allow the fluid to associate more effectively. For the designs analysed herein large areas of the active volume are free from particles, which possibly can be improved by going to smaller cell sizes. It is also possible that further improvements in the training procedures will allow to steer the ferrofluid more effectively.

**Figure captions:**

Fig. 1: Two interacting dipoles with magnetic moments **m**$_i$ and **m**$_j$, separated by distance **r**$_{ij}$.

Fig. 2: Magnetic thin film element used in the input and output pads, with the magnetic easy axis along the longer dimension due to shape anisotropy, and the magnetization (**M**) directed at angle θ with respect to the easy axis.

Fig. 3: The easy (a) and hard (b) axis magnetization loops of the pad element of Fig. 2 as predicted by the single domain theory. The bi-stability at zero field of the easy axis loop is used for magnetically inputing digital patterns into the ferrofluid layer, with digital 0 and 1 corresponding to the magnetization of the pad element up or down.

Fig. 4: An illustration of the field profile produced by the array of input/output pads during training – the bright areas correspond to regions of high field and dark areas to regions of low field. The black rectangles represent the pads arranged in a 2D matrix.

Fig. 5: Cross section of an input pad. The input pad is in the parallel state during training for achieving high fringing field. It is switched into the anti-parallel state during recall to minimize the direct influence on the output pads. The top magnetic film is somewhat thicker than the bottom film to avoid a completely flux-closed state during recall. The top layer is taken to be 25 nm thick, while the bottom layer is 20 nm thick  The metallic non-magnetic spacer is 10 nm thick.

Fig. 6: Cross section of an output pad. The output pads are in the parallel magnetic state during training, similar to the input pads. The top and bottom magnetic layers have the same thickness, 20 nm, to avoid any fringing fields from the output pads in the anti-parallel magnetic state during recall. The spacer is 10 nm thick. The output  is used as a field sensor during recall, owing to its classical spin-valve structure.

Fig. 7: Time trace of the current through an output pad during readout. The soft magnetic layer will switch magnetization direction when the current reaches I$_0$, if there is no field from the ferrofluid. The field from the ferrofluid will either increase or decrease the field sensed by the pad, thereby shifting the switching field left or right. This change in the switching field is determined by measuring the resistance of the tri-layer spin-valve pad.

Fig. 8: Particle configuration in the ferrofluid layer after 100,000 MC steps of simulation using the Metropolis algorithm without cluster-moving (a) and with cluster-moving (b). A constant magnetic field is applied in the horizontal direction for this test simulation. The cluster-moving algorithm produces particle patterns with a higher degree of aggregation.

Fig. 9: Definitions for computing the field from a thin uniformly magnetized magnetic element.

Fig. 10: (a) Illustration of the undesirable zero-link configuration – all four dipole interactions disfavor this bit pattern. (b) The opposite output pad state is highly favorable. This is a shortcoming of the standard dipole pad layout.

Fig. 11: Quadrupole input pad layout used to eliminate the zero-link problem – all local bit input/output patterns are equally favorable.

Fig. 12: Typical outcome of using the alternating pattern training procedure with the quadrupole input pad layout of 3x3. All bits are recalled correctly on average, but large deviations can occur. The error bars correspond to one standard deviation in each direction. 20 independent recall attempts were made for each pattern.

Fig. 13: The configuration of the particles after training. The output pads are shown in red and the input pads in green. The input pads are 80 x 220 nm and the output pads are 350 x 80 nm. The footprint of one cell is 490 x 620 nm.

Fig. 14: Large scale simulation. Two pattern pairs to be stored by the ferrofluid memory device: (a) pattern 1 and (b) pattern 2. The patterns are selected such that at least one link configuration is different for each bit between the two patterns. Simultaneous association of such two patterns results in conservation estimates of the performance of the device.

Fig. 15: The difference between the two patterns selected for training. Shaded icons indicate that the output is different and the numbers show how many of the four closest input bits change.

Fig. 16: Design improvement consisting in arranging the elongated output pads in different directions, which minimizes the interference between the output pads and creates essentially the same magnetic environment for all bits. The input pads are 60 x 180 nm, the output pads 60 x 300 nm, and the cell size 300 x 300 nm.

Fig. 17: The configuration of particles after every one hundred thousand MC steps of alternating-pattern training was saved and the recall accuracy was probed. Since pattern 2 was the last in the cycle, the results are biased in favor of this pattern, where the recall accuracy approaches 100%. The recall accuracy of pattern 1 is ~80%.

Fig. 18: Recall accuracy with pattern 1 stored at 600 K and pattern 2 stored at 300 K as a function of the readout threshold field setting.

Fig. 19: The particle configuration after (a) only one pattern was trained at 600 K, and (b) two patterns were trained, one at 600 K and the other at 300 K. The two particle distributions are significantly different. Configuration (b) yields 90-100% recall of both patterns.

Fig. 20: Summary of the results for the two training algorithms: variable-temperature (top) and alternating-pattern (bottom). The latter algorithm favors the last pattern trained. The color distributions in the middle section are the field maps from the ferrofluid dipoles as sensed at the output pads, after the fluid settled magnetically during recall. The field maps for the two patterns are significantly different, which illustrates the core action of the proposed nano-magnetic network based memory – the actual memory consists of spin distributions that is dynamic in nature, realized only in response to the input patterns that the system has been trained for.

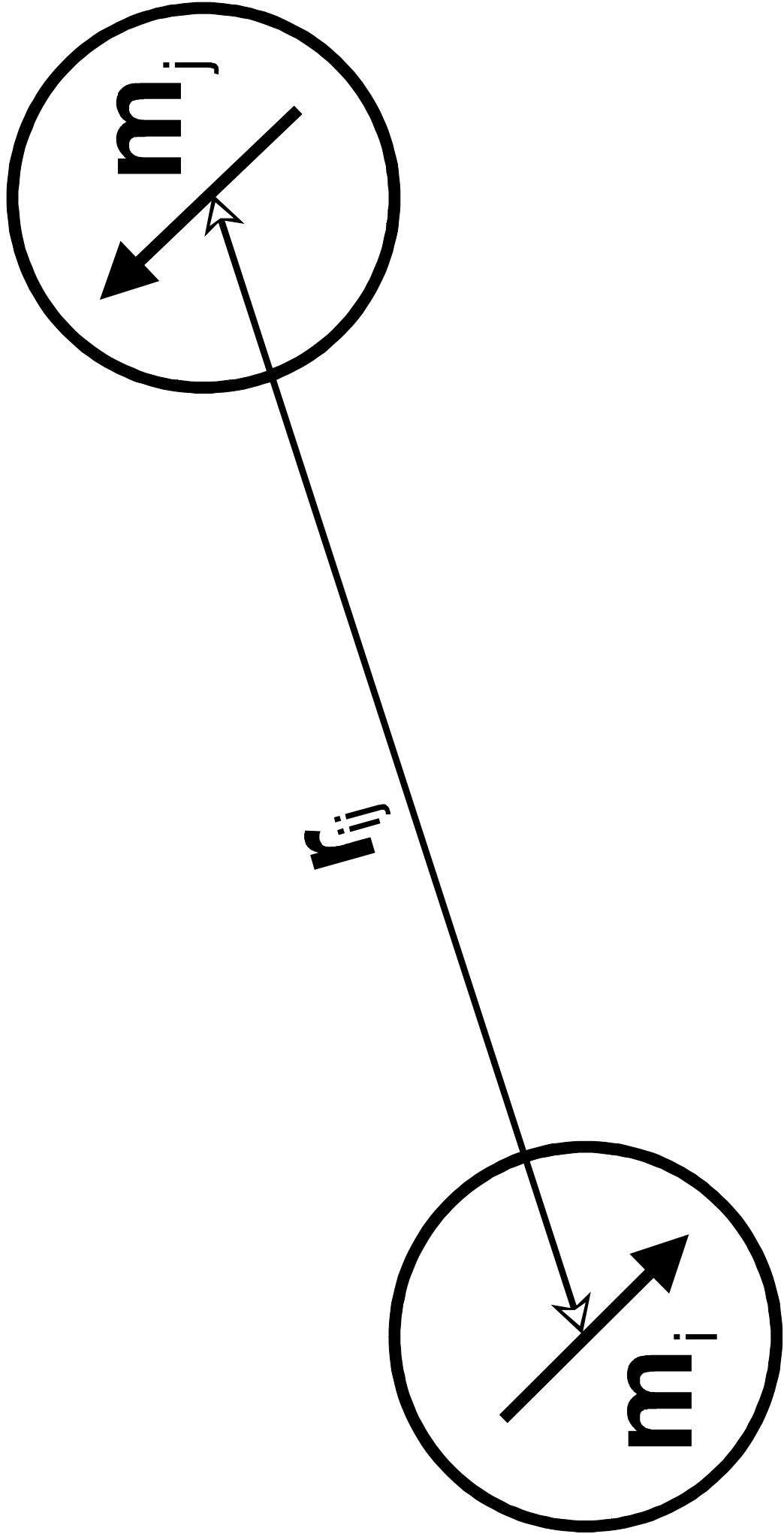

Figure 1

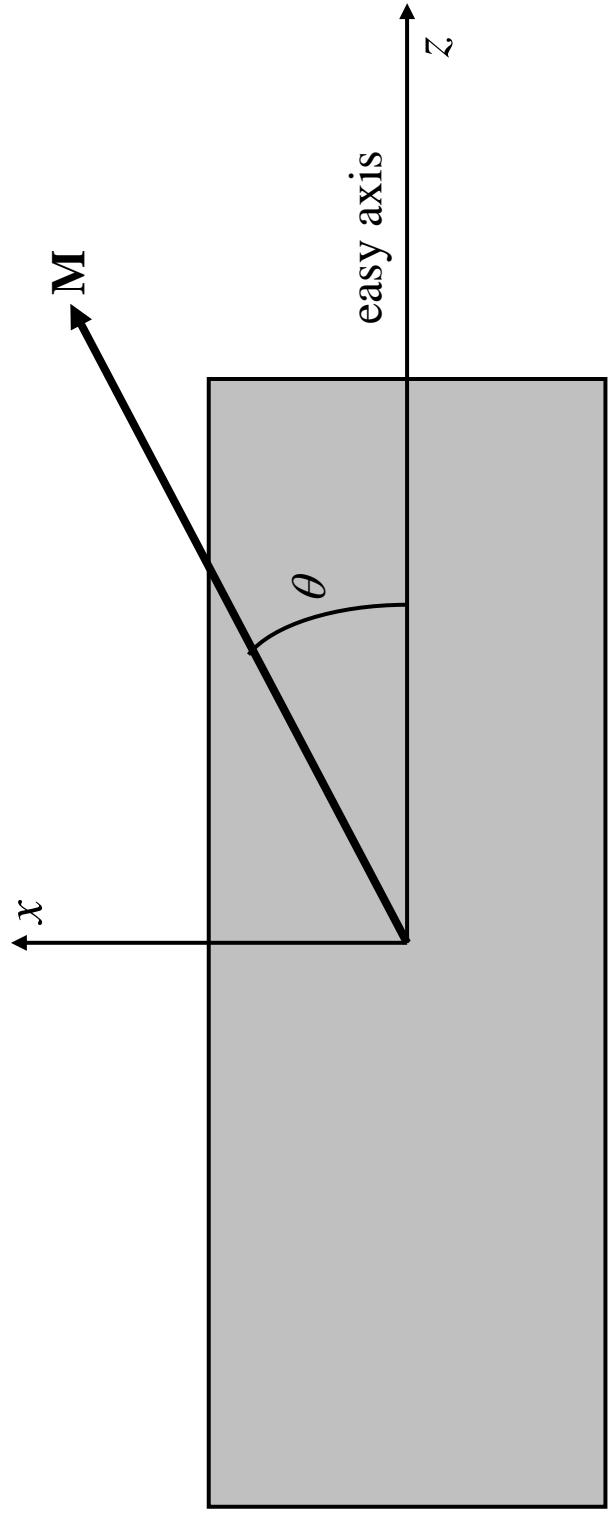

Figure 2

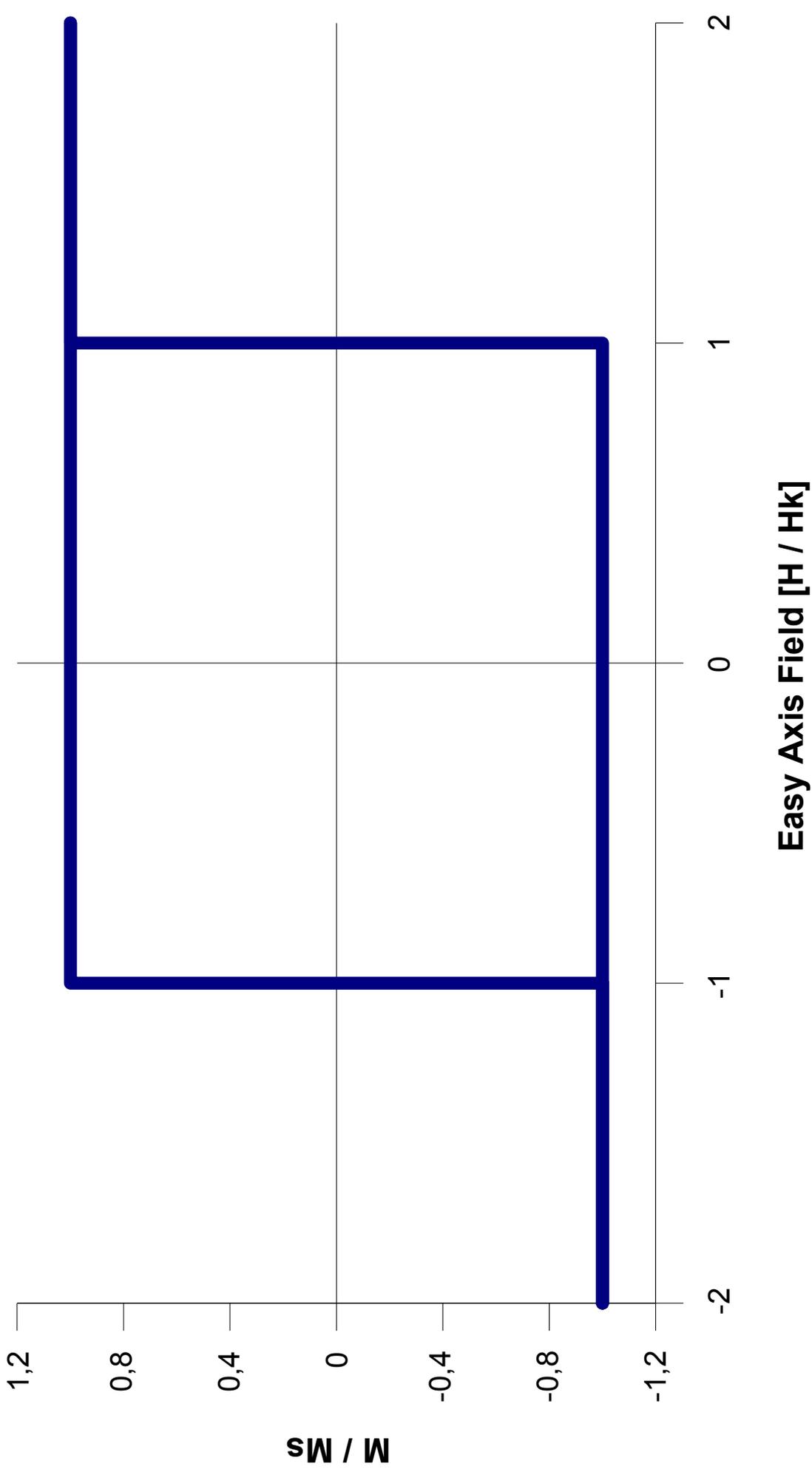

Figure 3

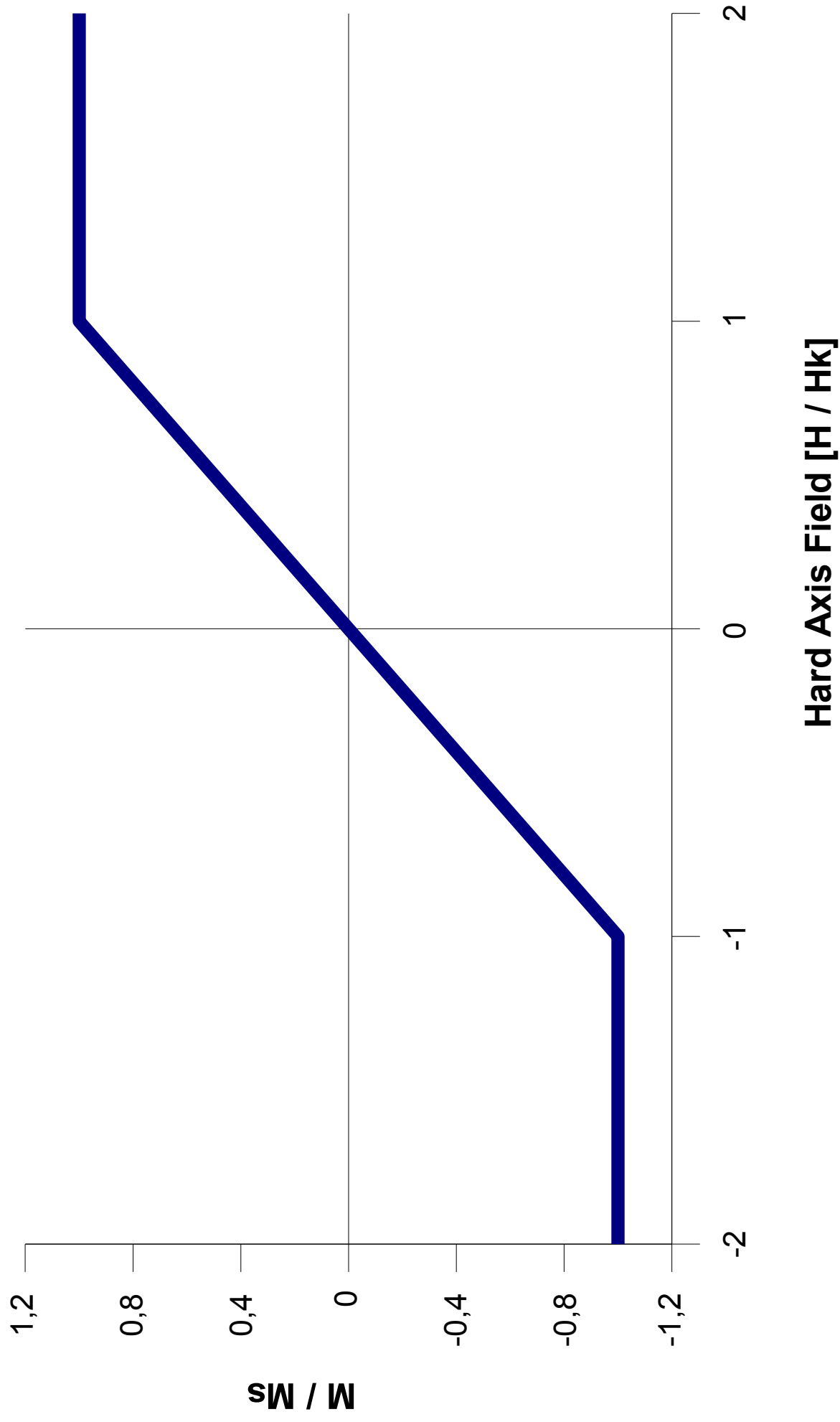

Figure 3

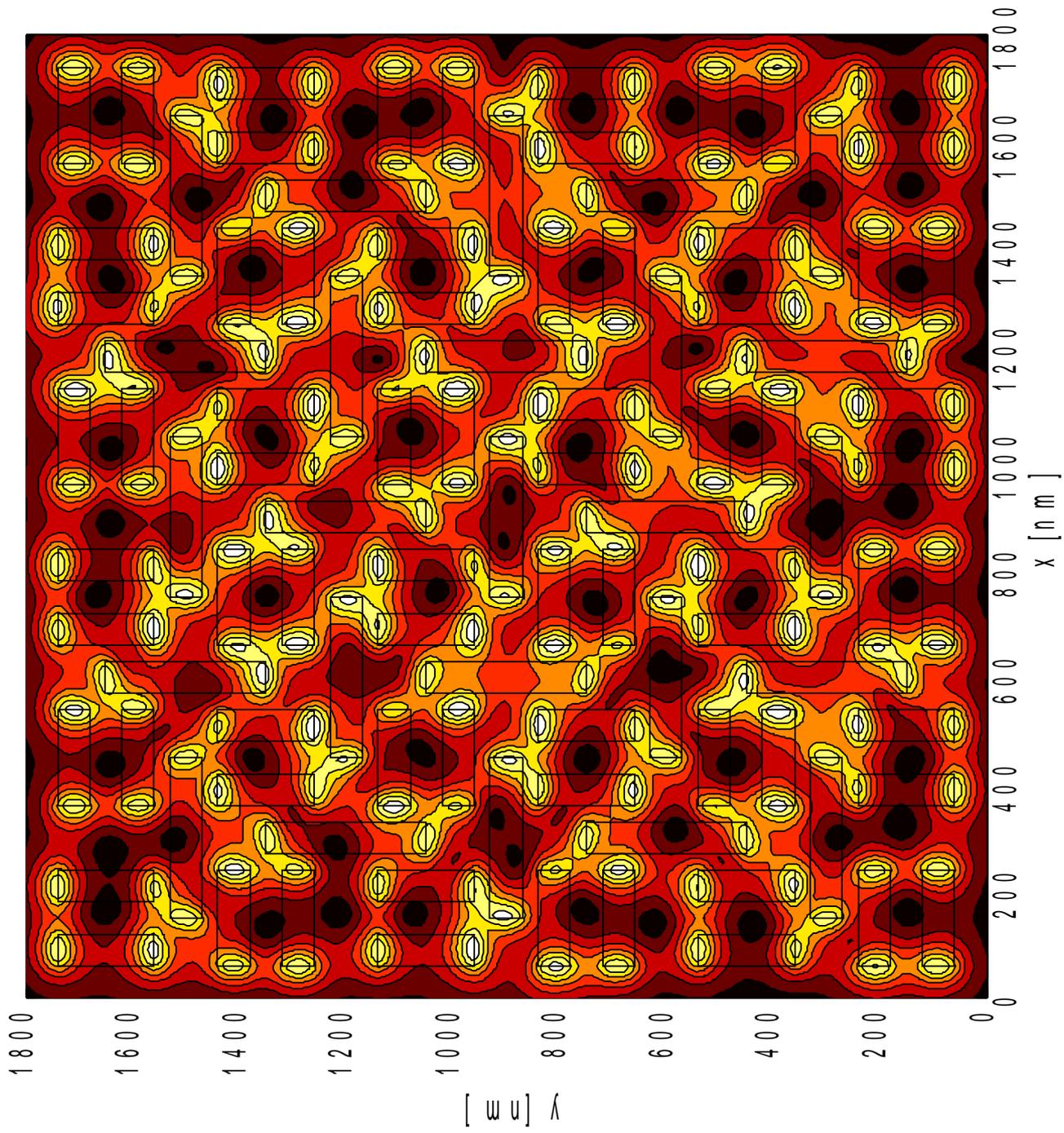

Figure 4

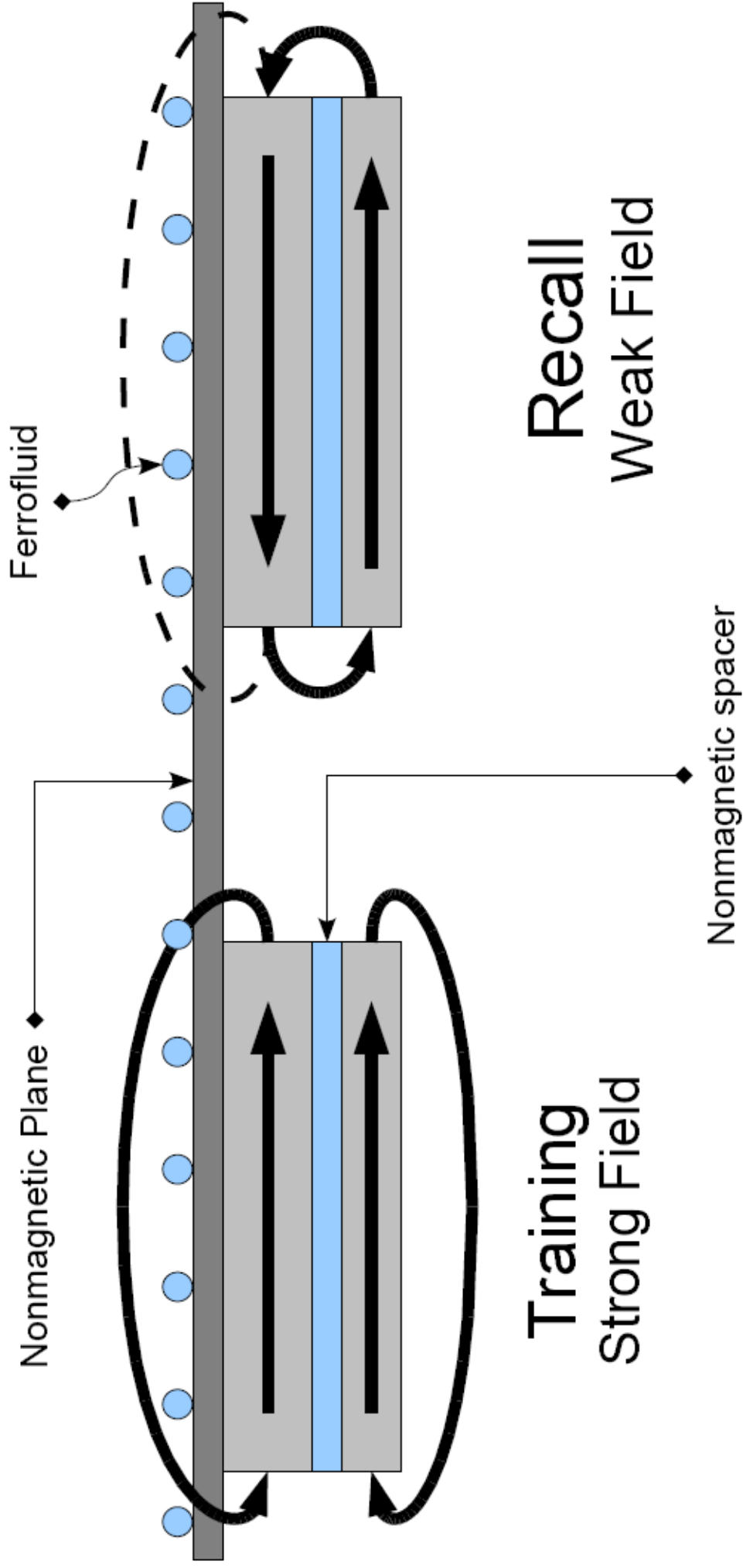

Figure 5

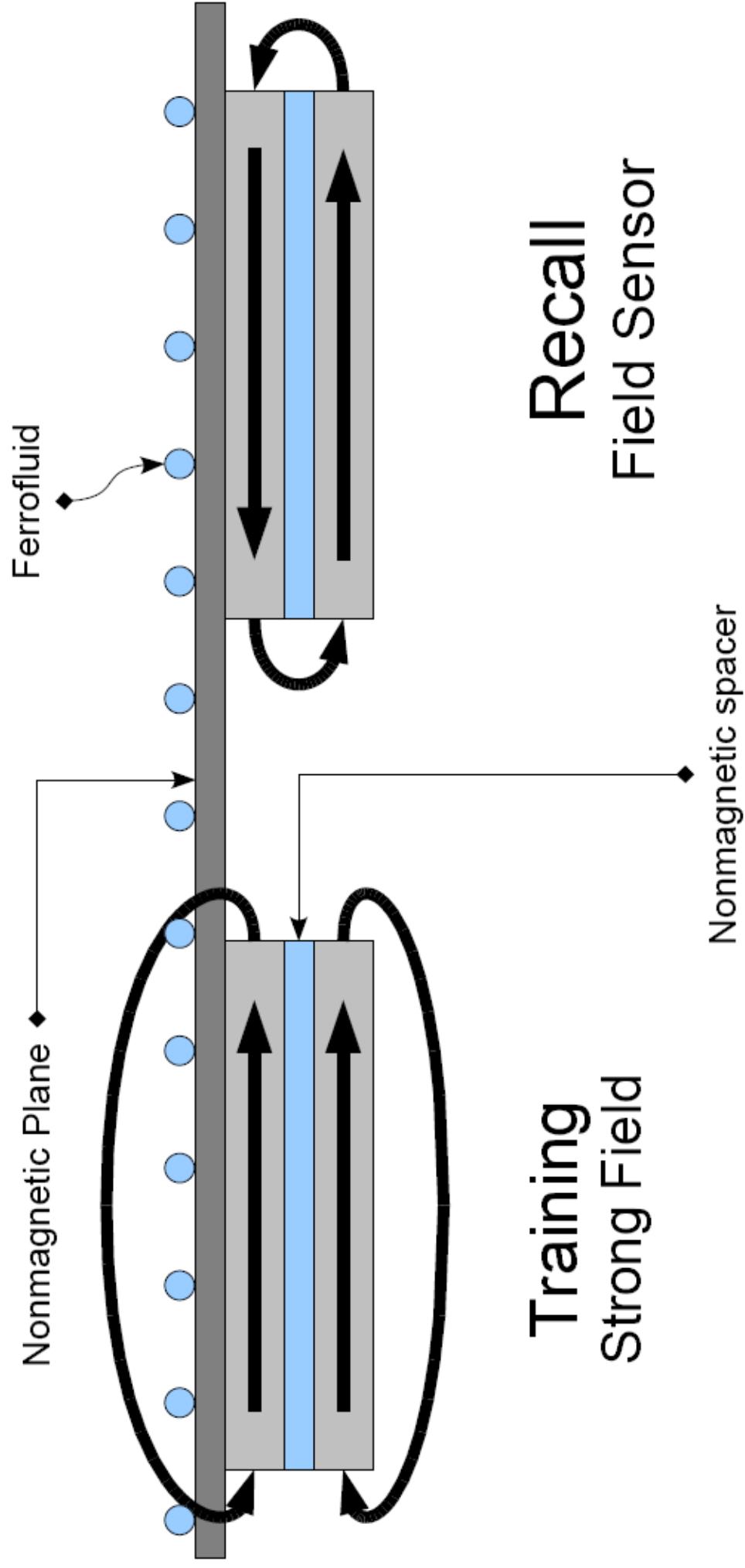

Figure 6

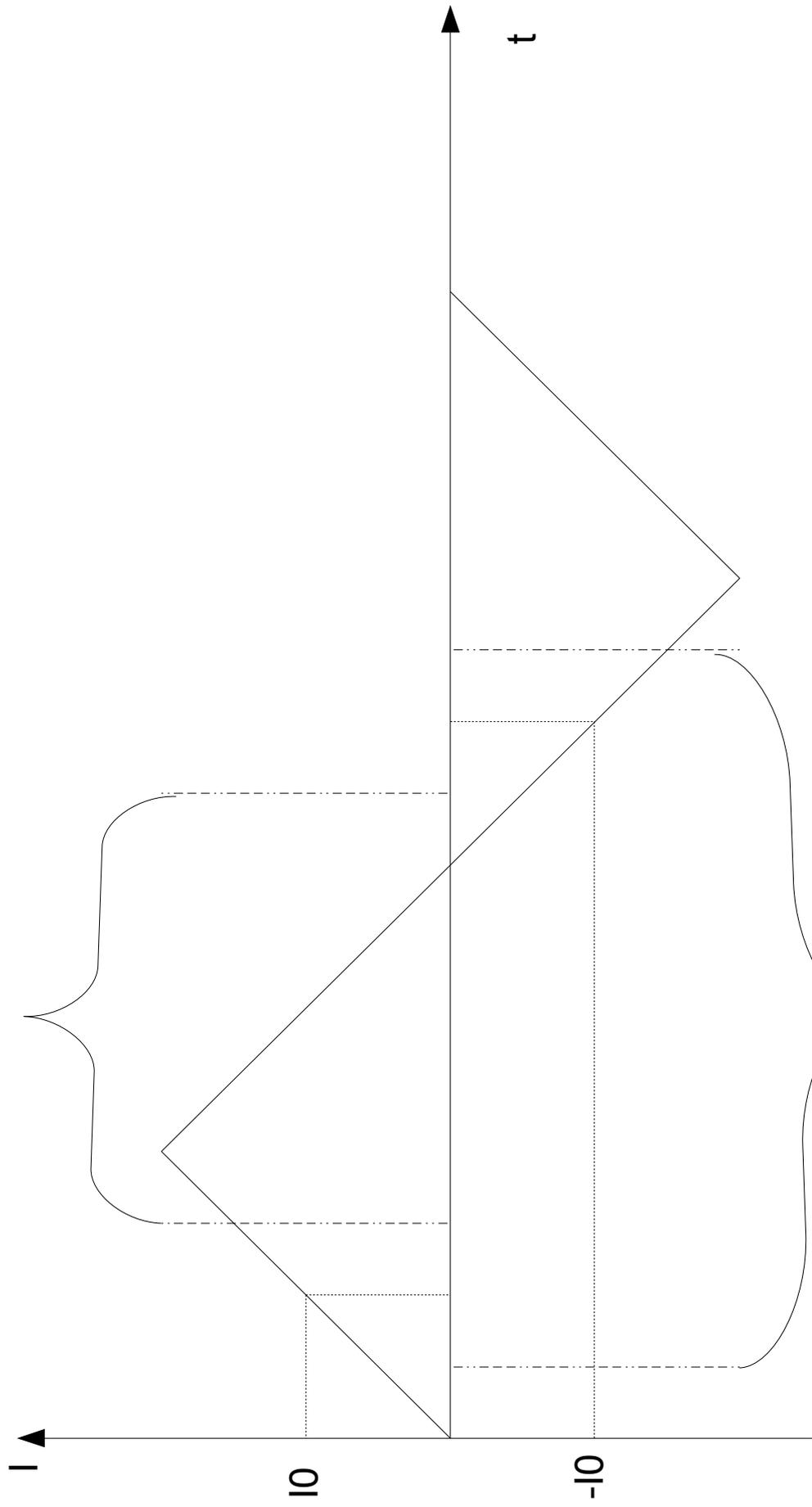

Figure 7

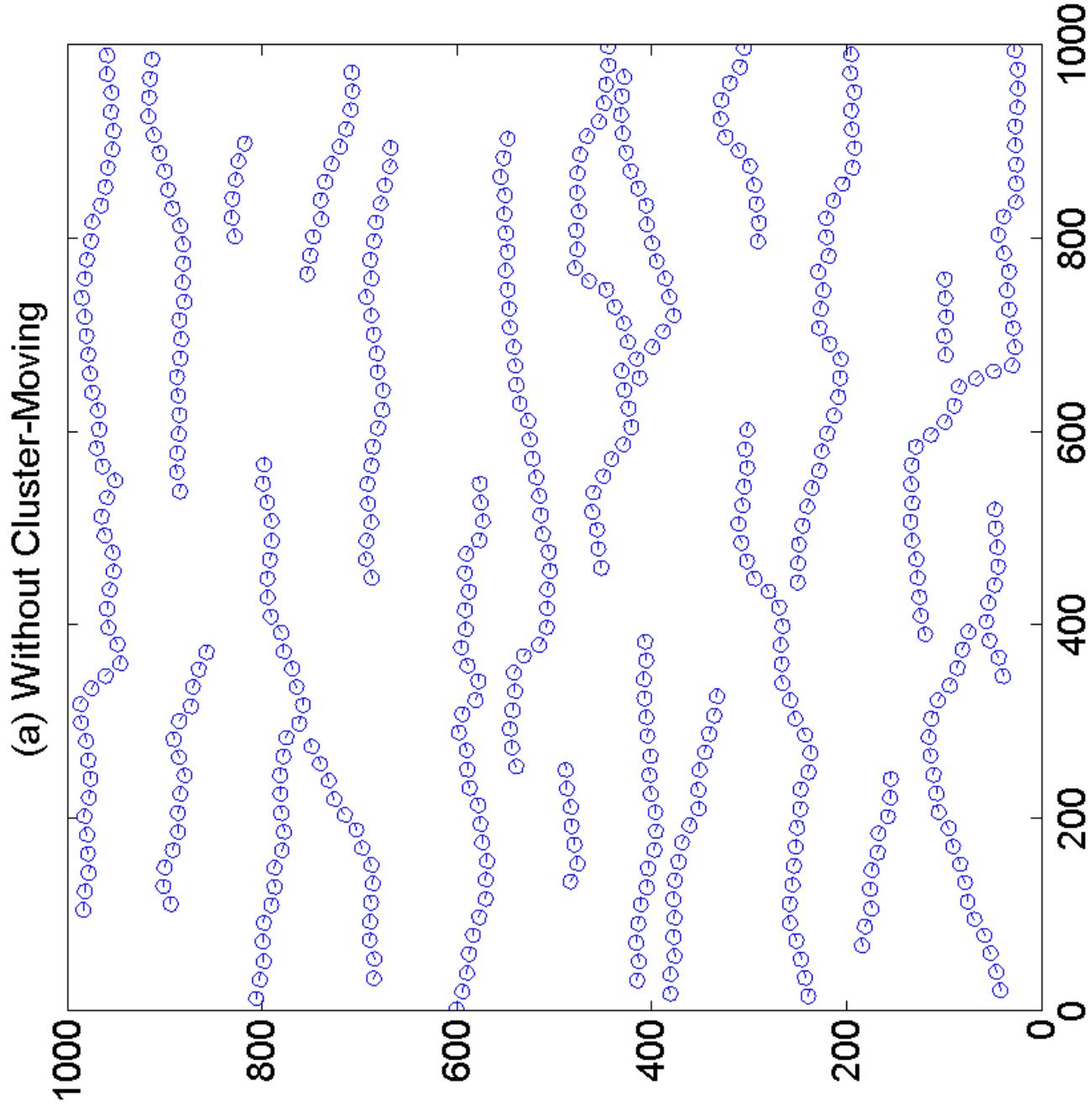

Figure 8

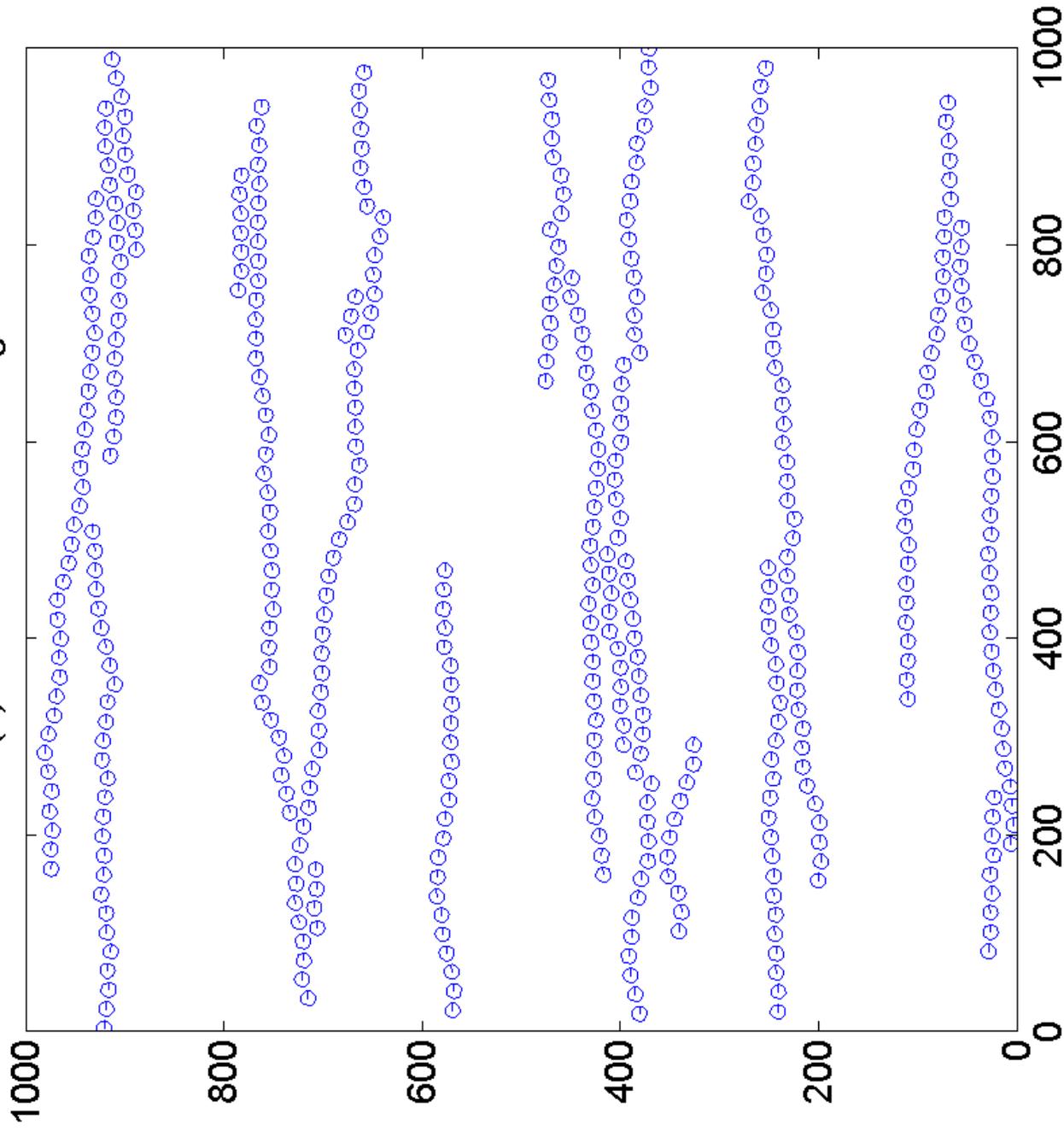
(b) With Cluster-Moving

Figure 8

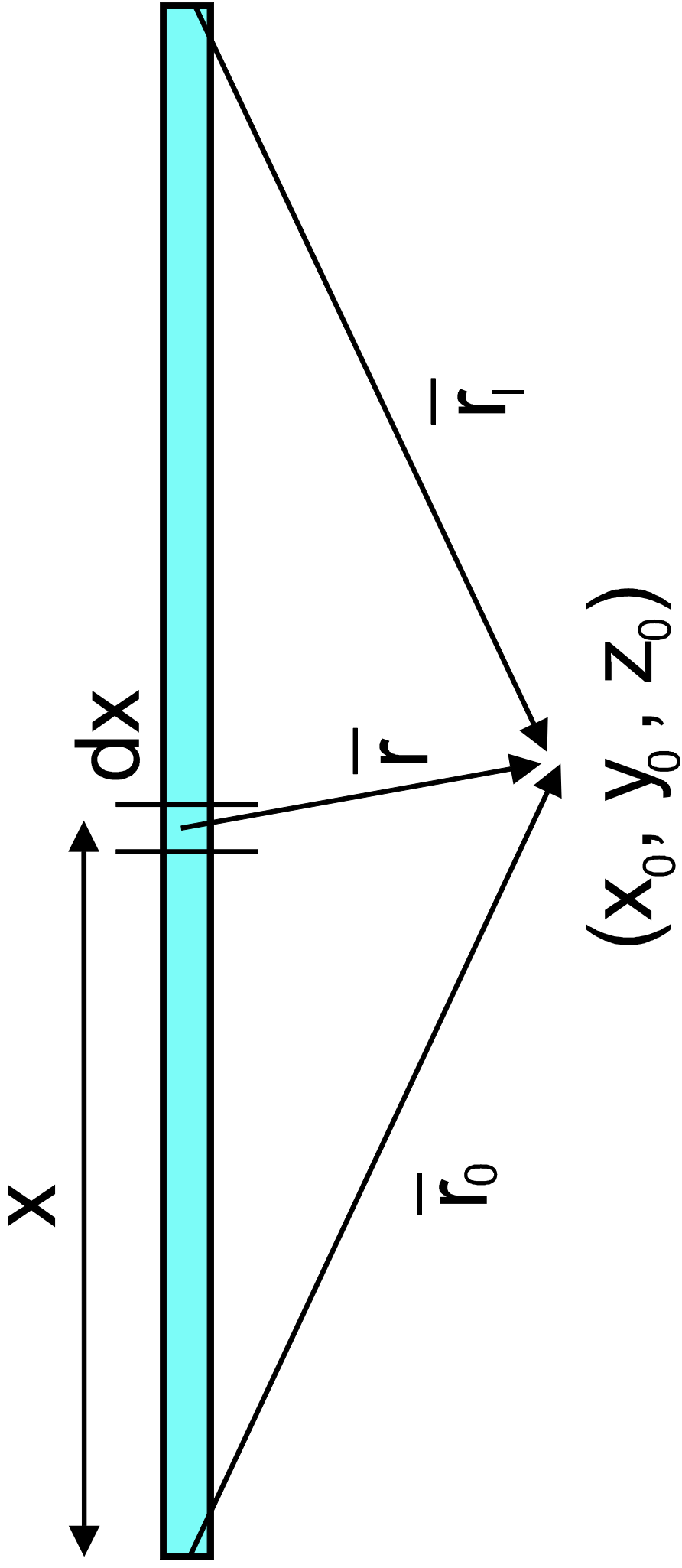

Figure 9

(a) zero link

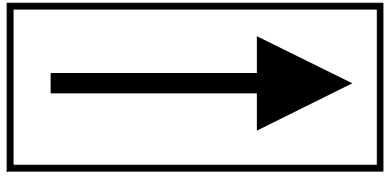
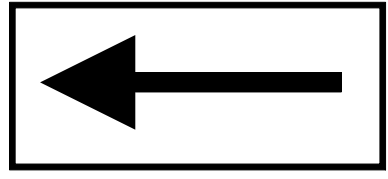
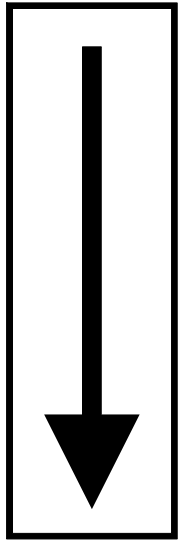
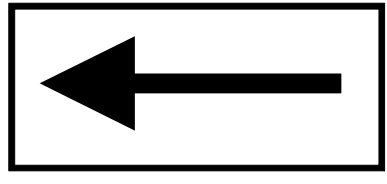
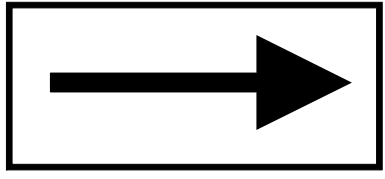

(b) four links

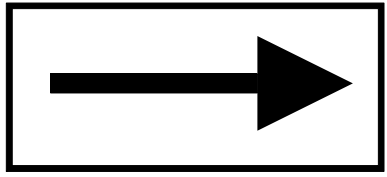
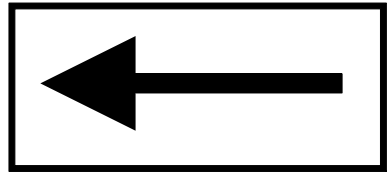
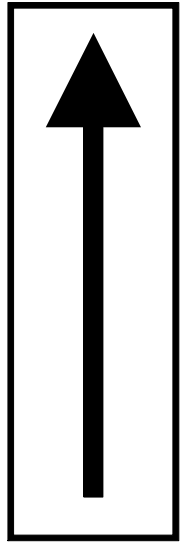
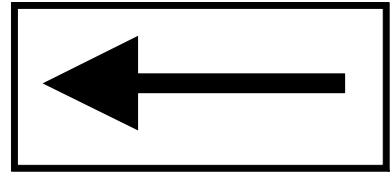
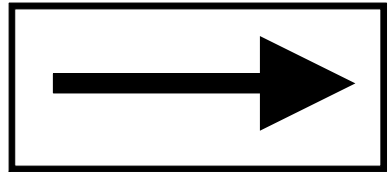

Figure 10

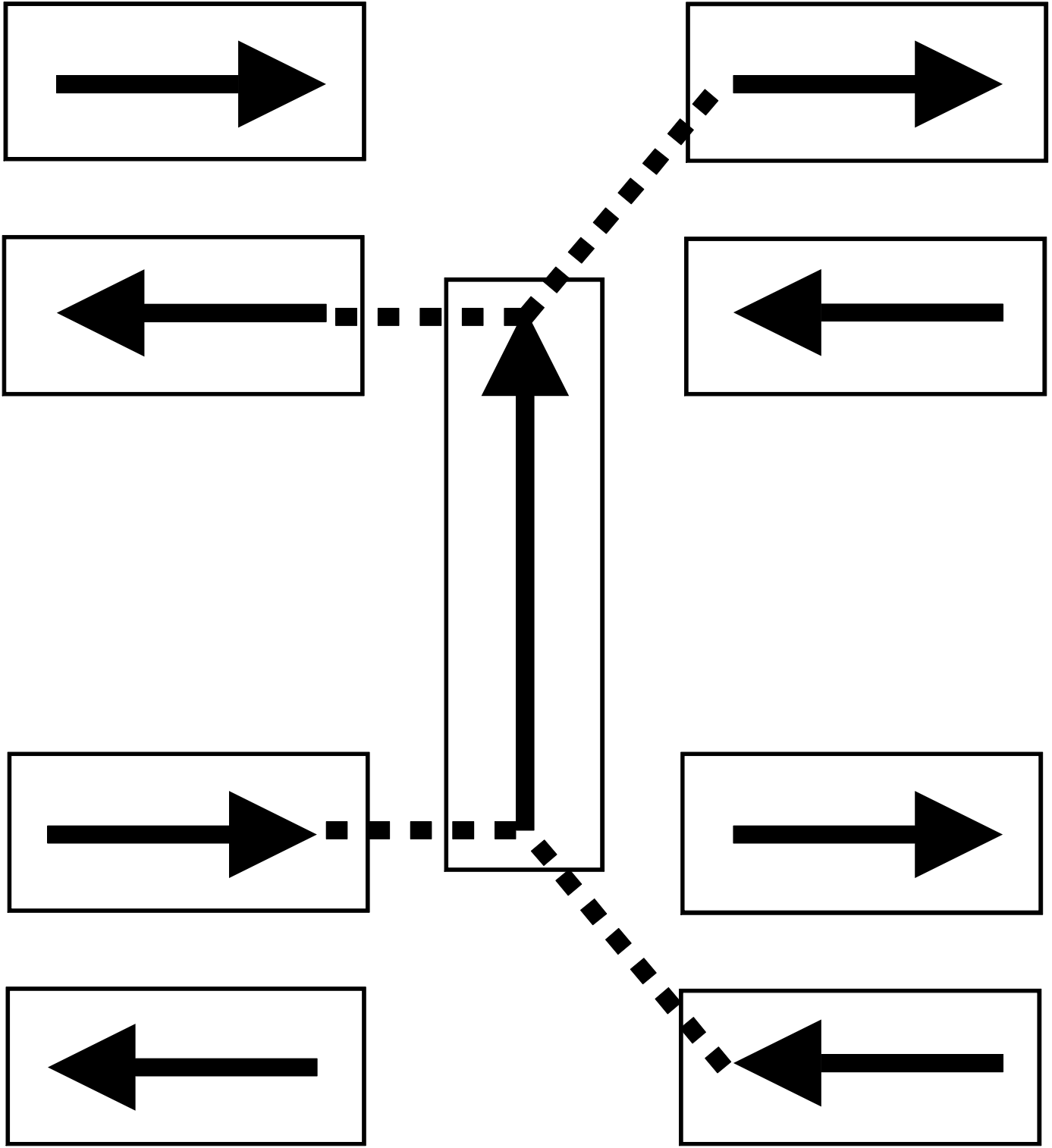

Figure 11

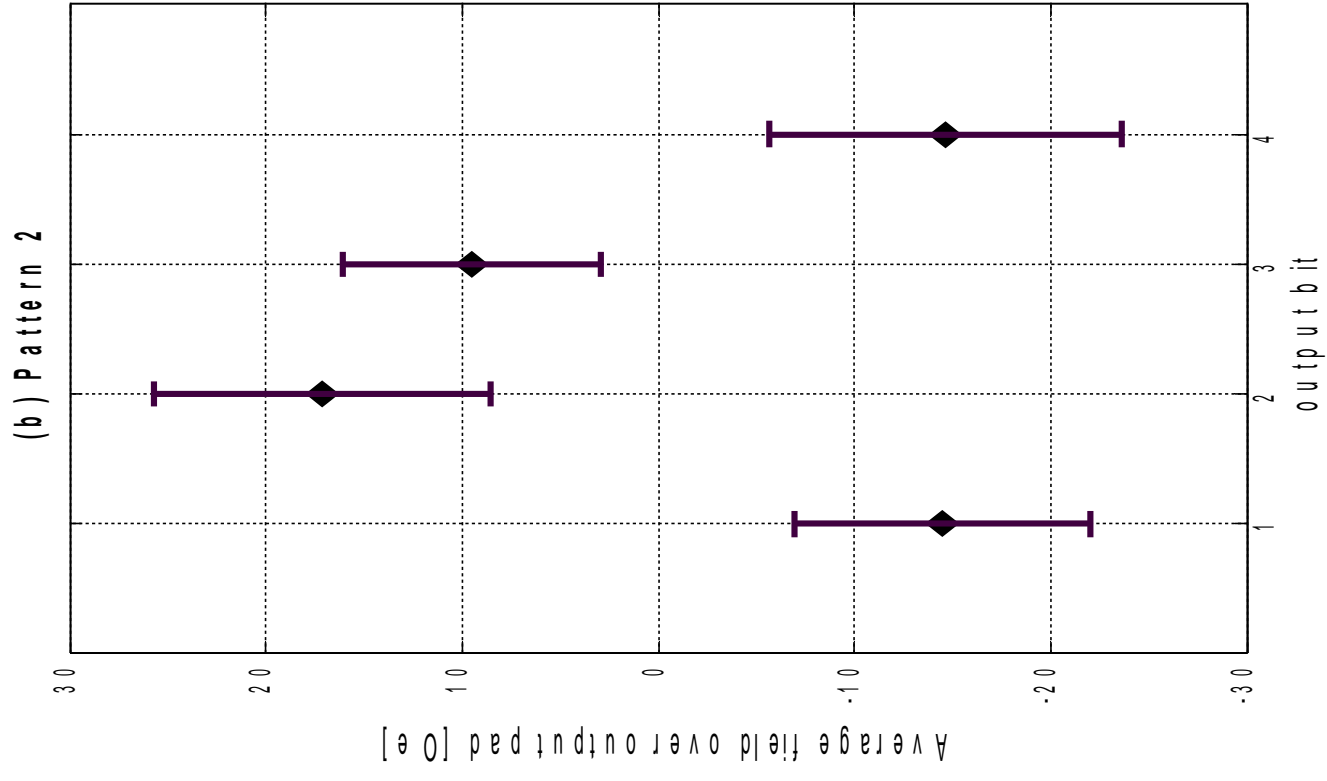

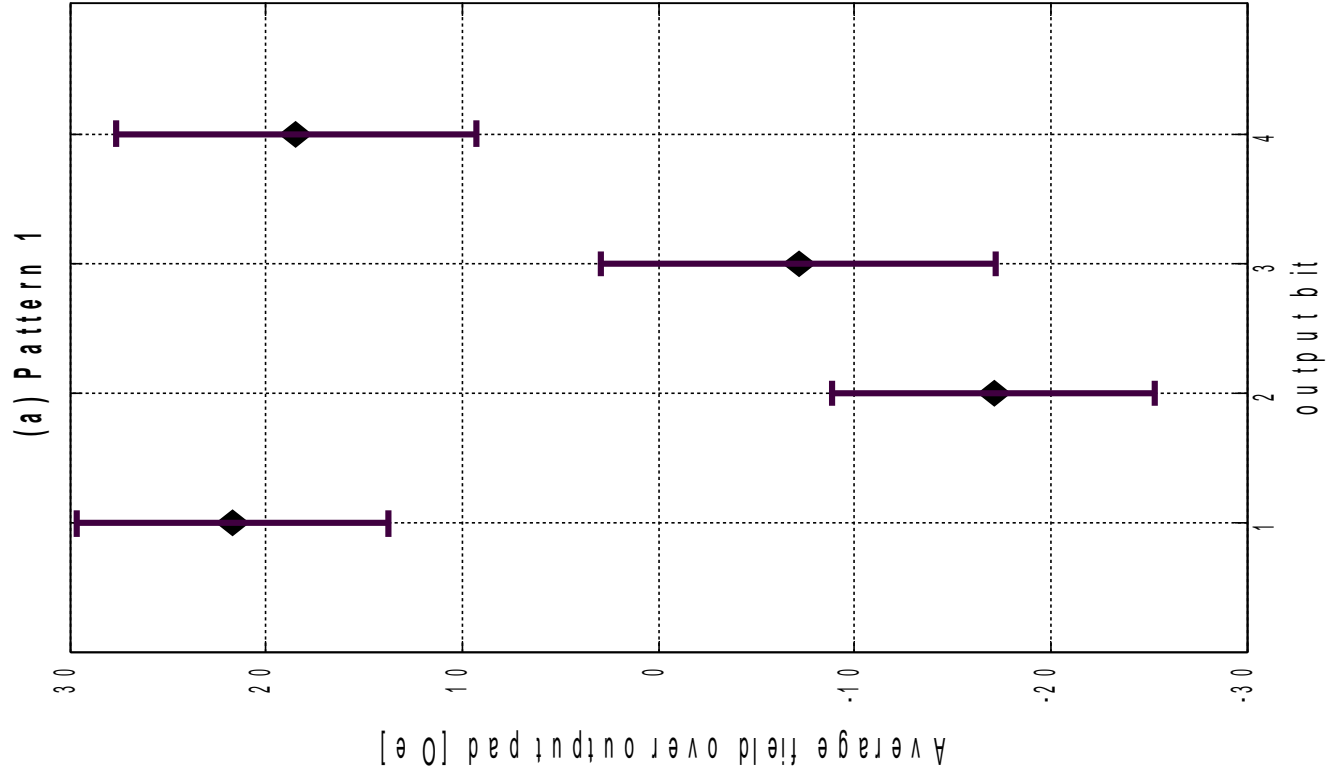

Figure 12

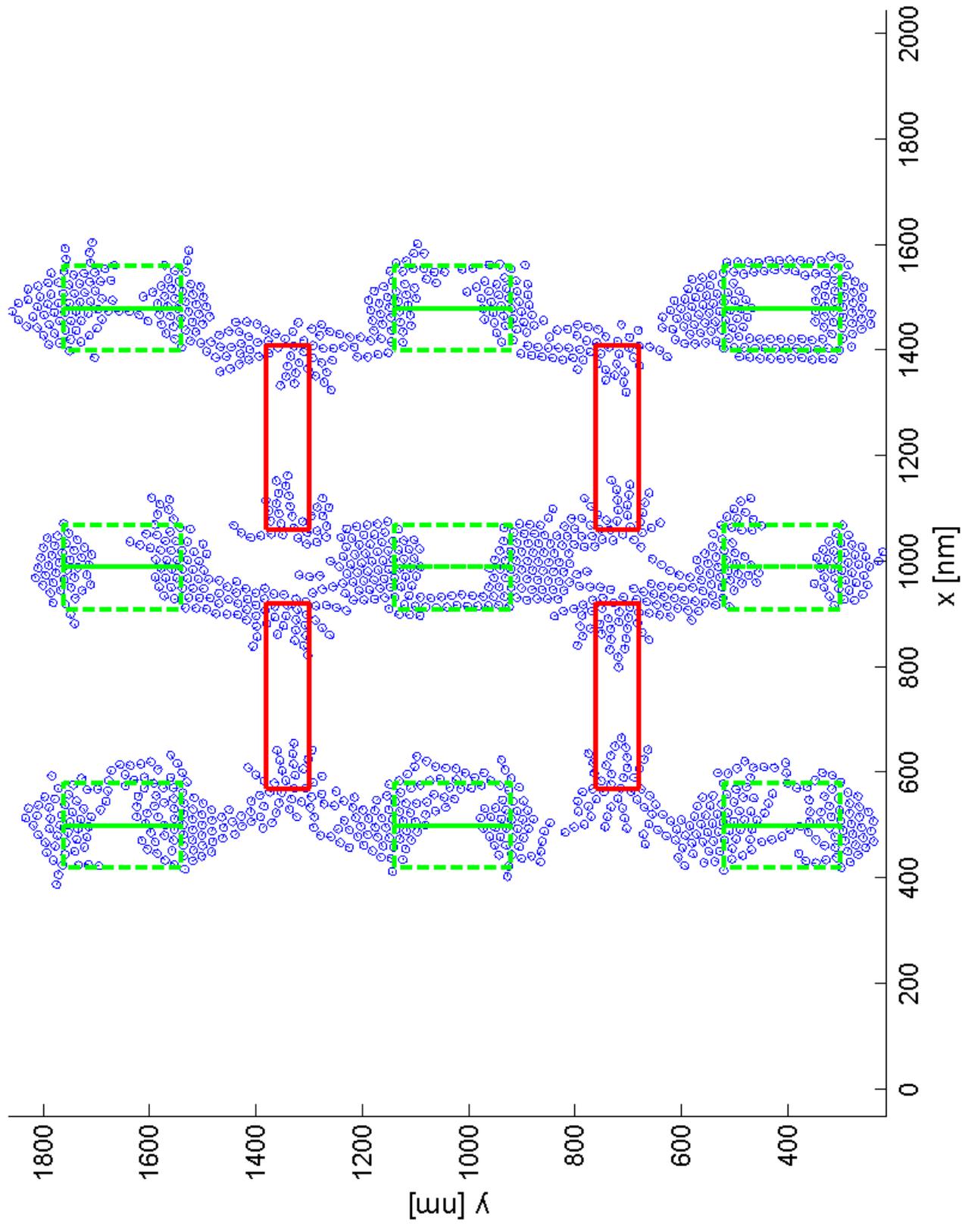

Figure 13

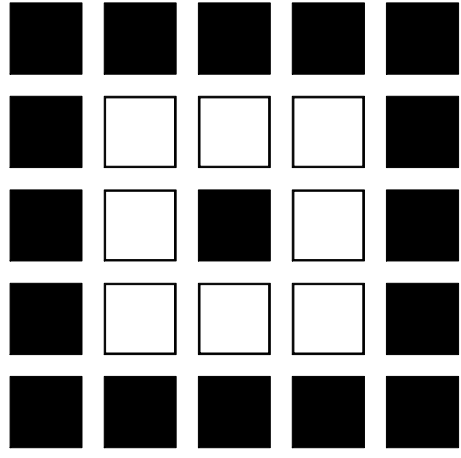
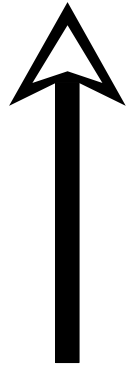
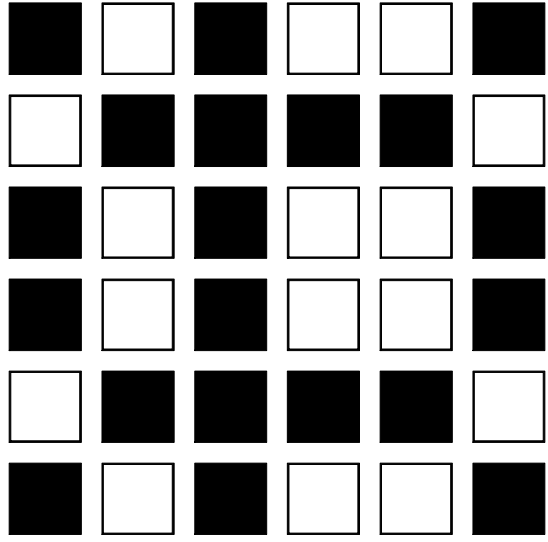

(a)

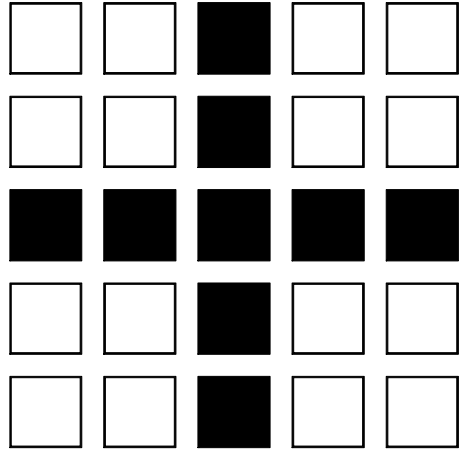
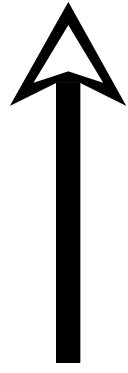
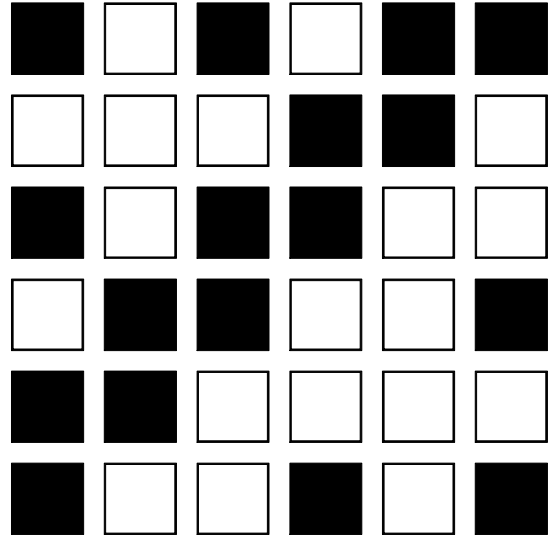

(b)

Figure 14

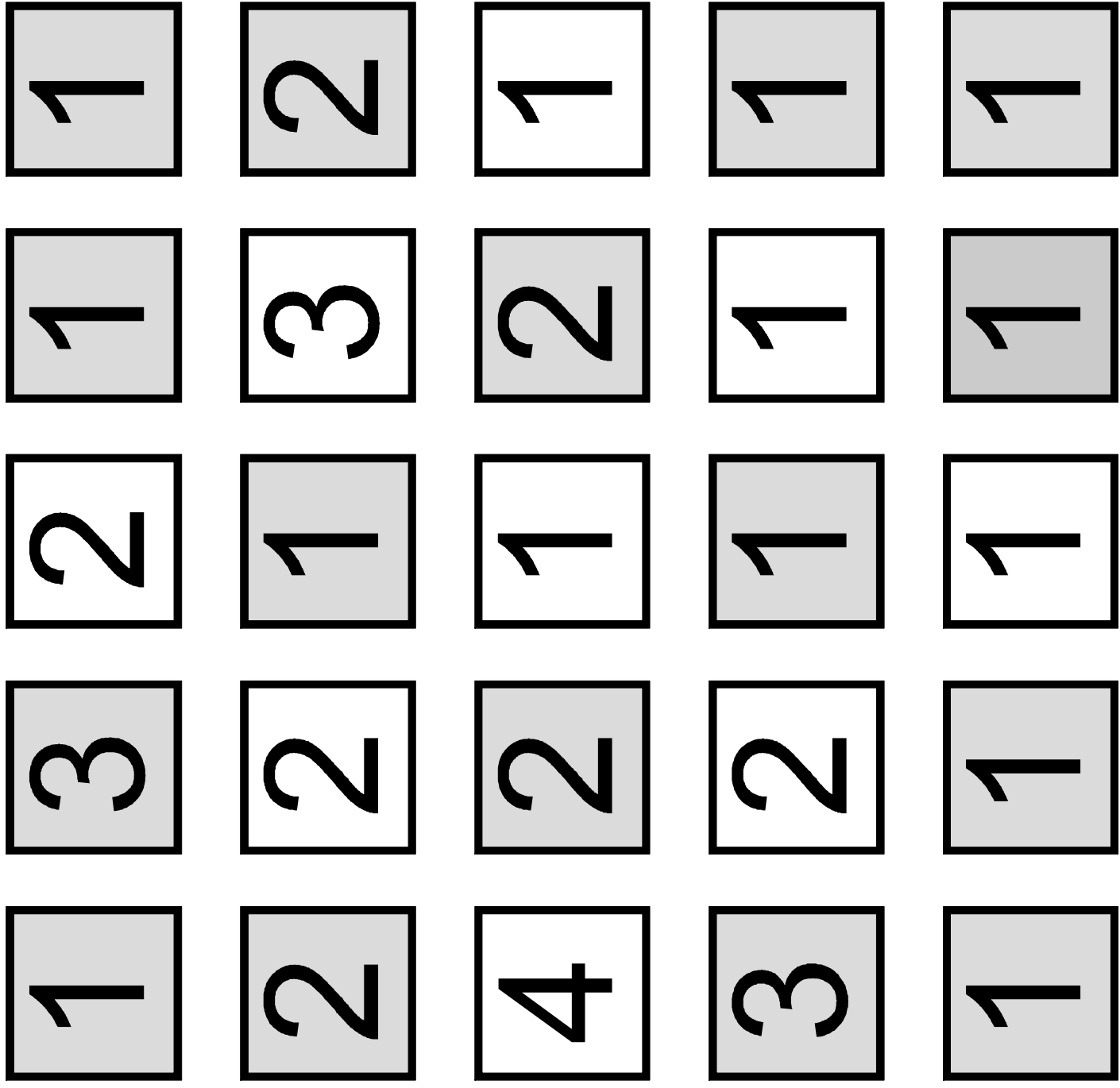

Figure 15

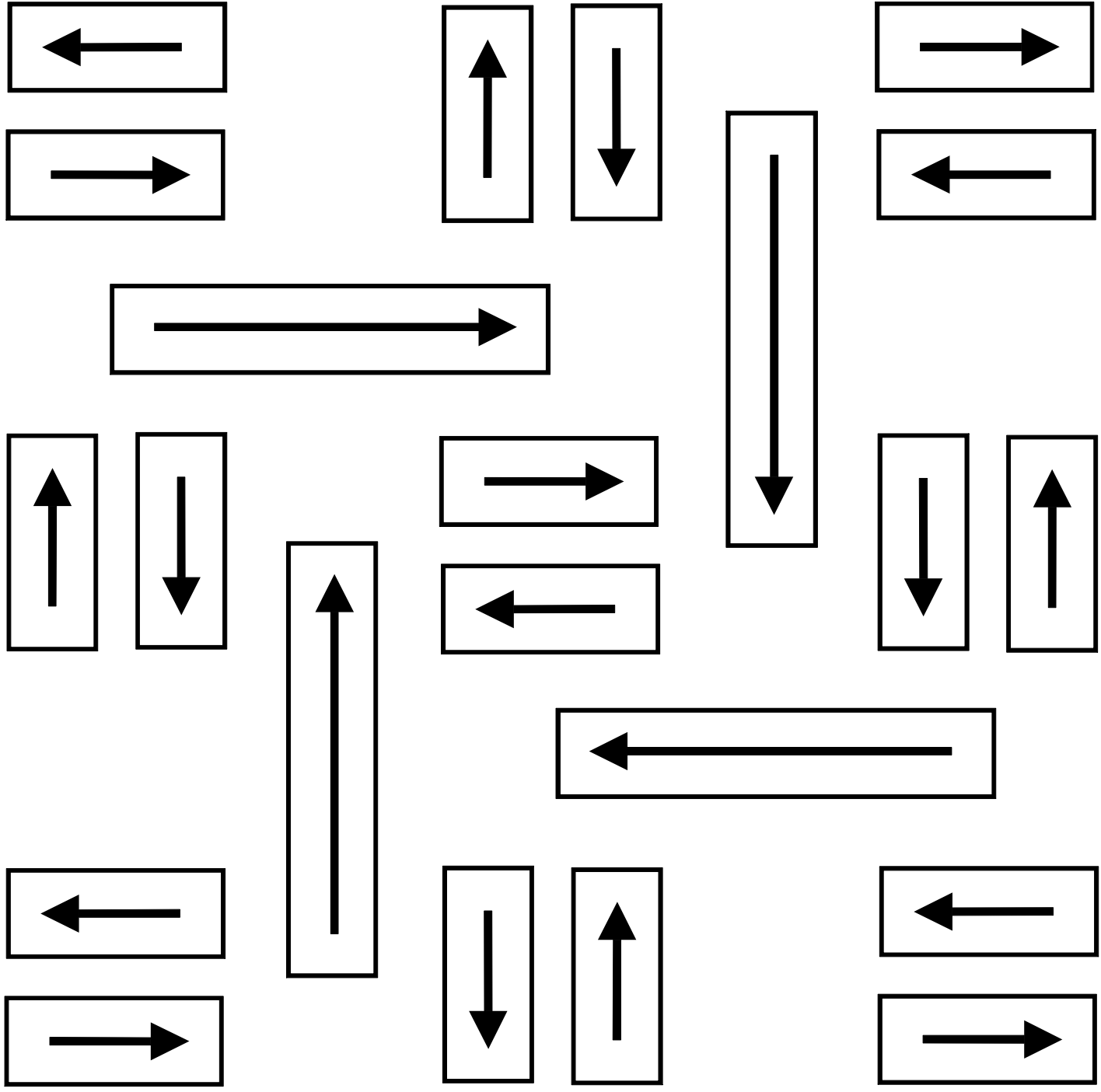

Figure 16

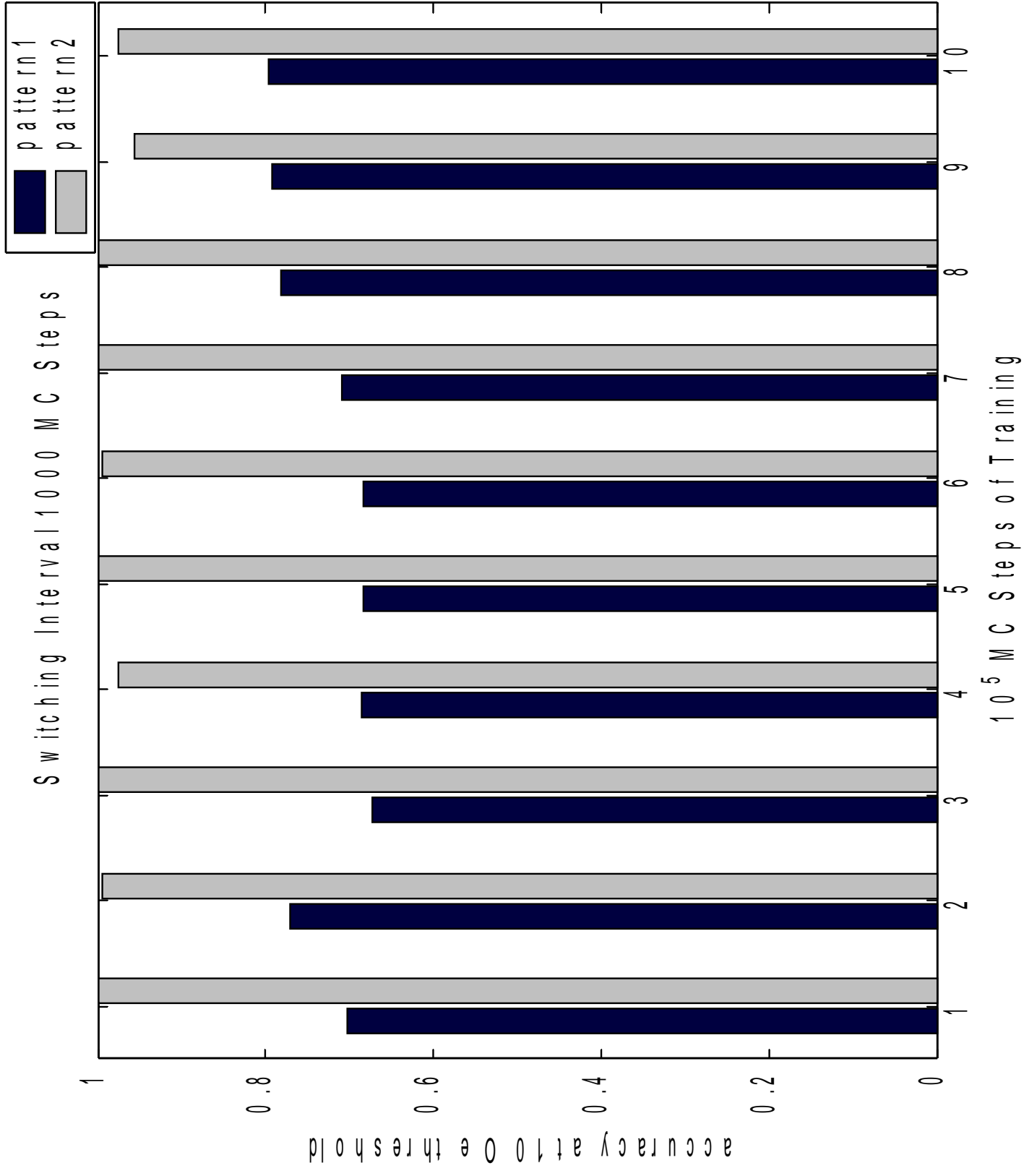

Figure 17

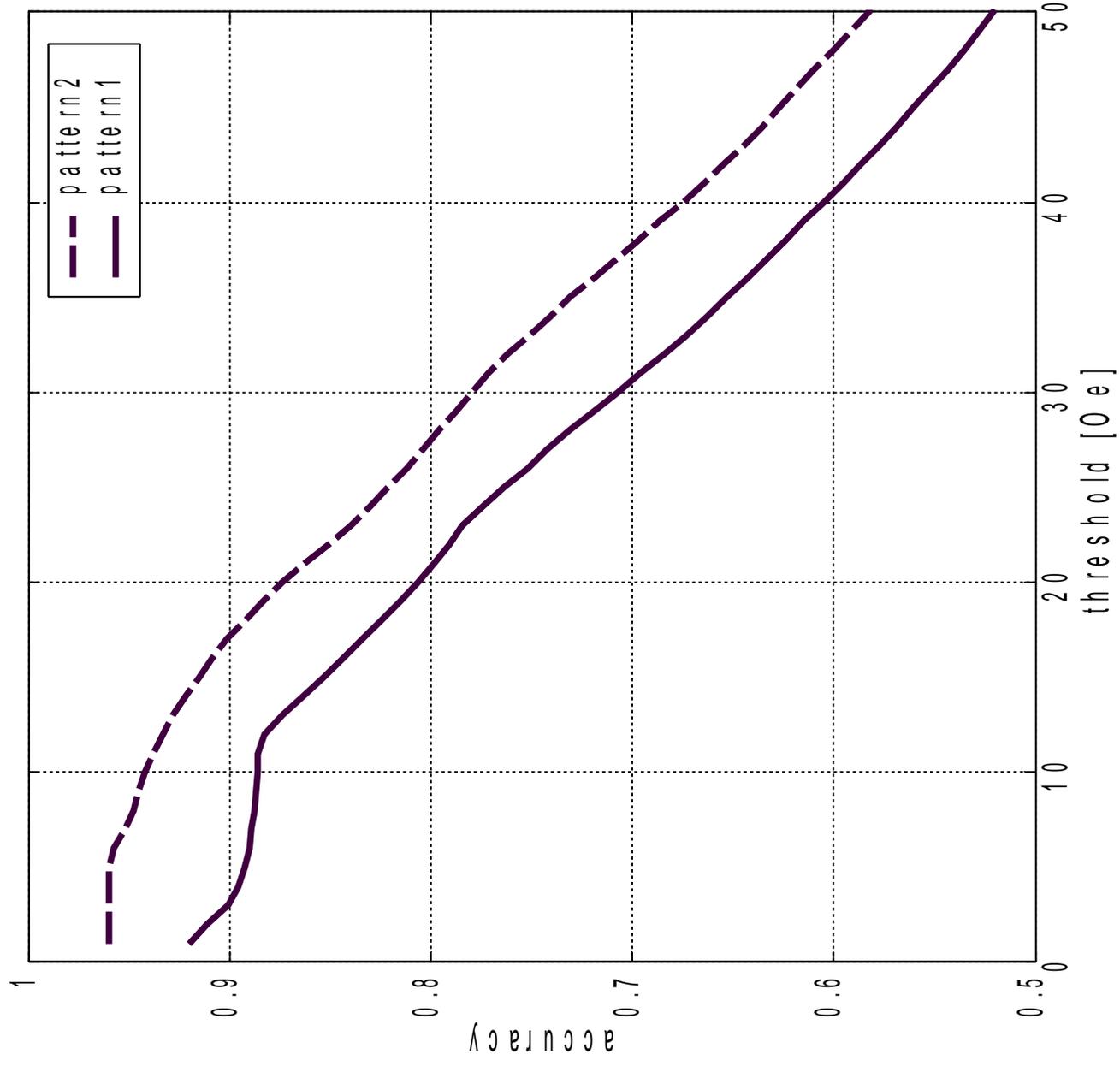

Figure 18

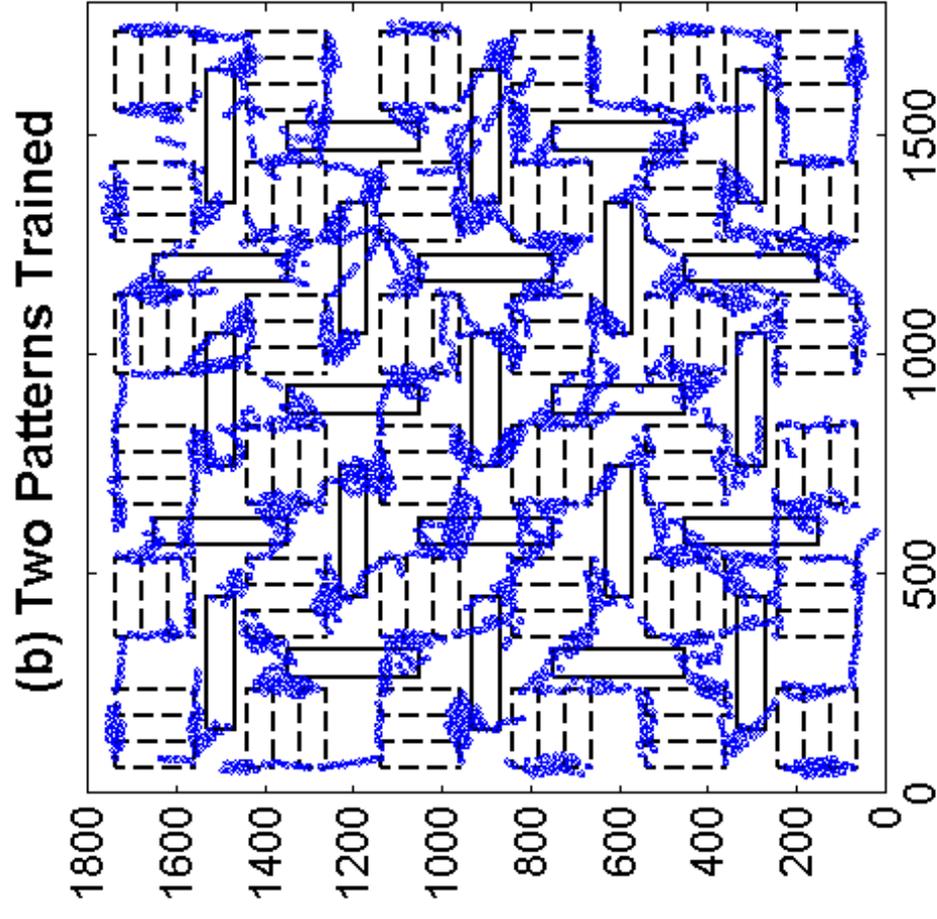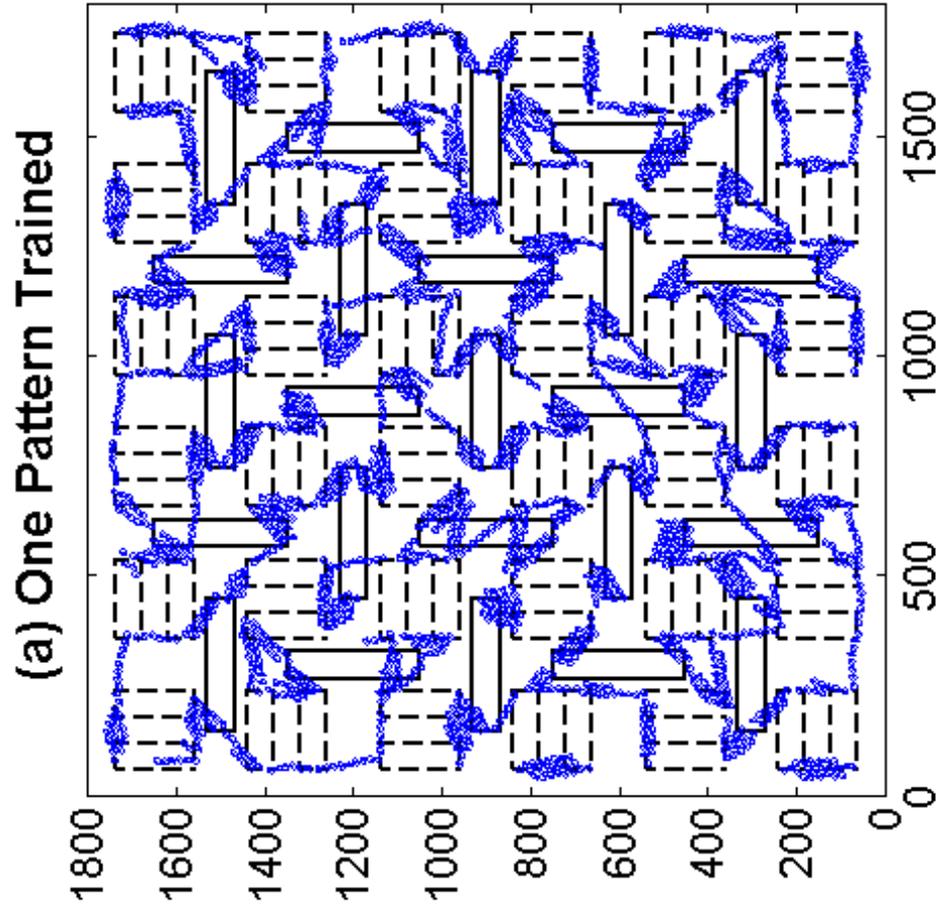

Figure 19

## Patterns stored at different temperatures

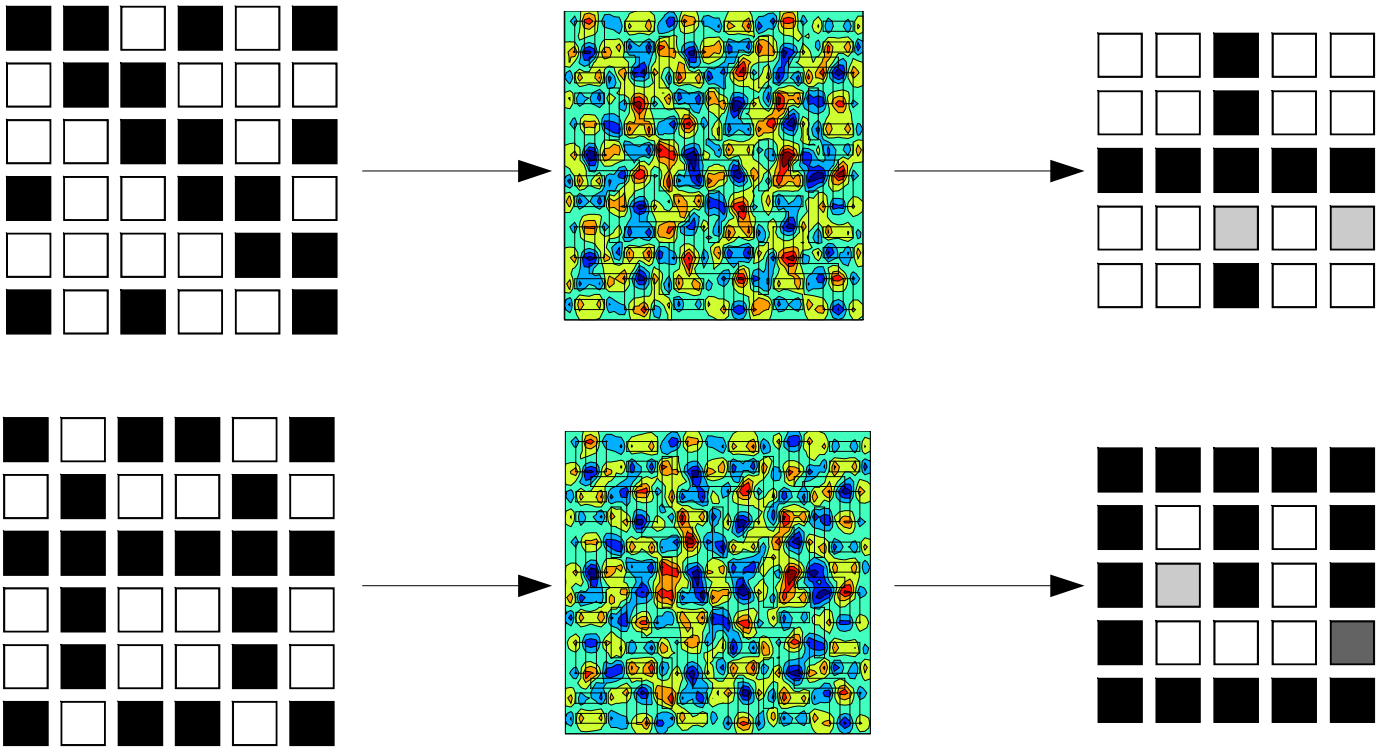

## Switching between patterns during training

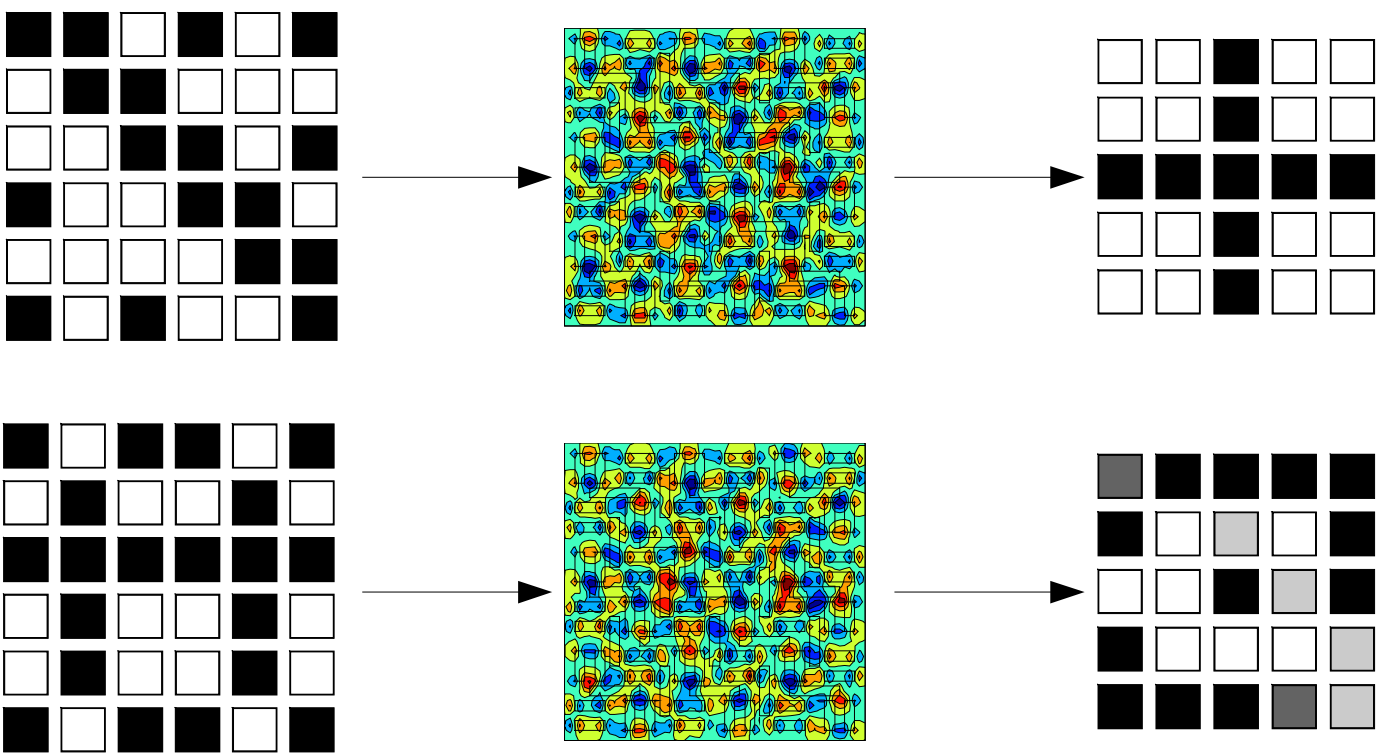

Figure 20